\documentclass[fleqn,usenatbib]{mnras}

\usepackage{newtxtext,newtxmath}

\usepackage[T1]{fontenc}
\usepackage{ae,aecompl}

\usepackage{graphicx}	
\usepackage{amsmath}	
\usepackage{color}
\usepackage{soul}
\usepackage{threeparttable}

\newcommand{\gppr}{\stackrel{>}{\scriptstyle \sim}}
\newcommand{\gappr}{\raisebox{-0.4ex}{$\gppr$}}
\newcommand{\lppr}{\stackrel{<}{\scriptstyle \sim}}
\newcommand{\lappr}{\raisebox{-0.4ex}{$\lppr$}}

\newcommand{\G}{{\it Gaia }}
\newcommand{\msun}{${\rm M}_{\sun}$}

\usepackage[x11names]{xcolor}

\usepackage{ulem}

\title[100\,pc WDs classification from {\it Gaia}-DR3 and the VO]{Spectral classification of the 100\,pc white dwarf population from {\it Gaia}-DR3 and the Virtual Observatory}

\author[F. M. Jim\'enez-Esteban et al.]{
F. M. Jim\'enez-Esteban$^{1}$\thanks{E-mail: fran.jimenez-esteban@cab.inta-csic.es}, S. Torres$^{2,3}$, A. Rebassa-Mansergas$^{2,3}$, P. Cruz$^{1}$,
\newauthor
 R. Murillo-Ojeda$^{1}$, E. Solano$^{1}$, C. Rodrigo$^{1}$, M. E. Camisassa$^{4}$
\\
\\
$^{1}$ Centro de Astrobiolog\'{\i}a (CAB), CSIC-INTA, Camino Bajo del Castillo s/n, campus ESAC, 28692, Villanueva de la Ca\~nada, Madrid, Spain\\
$^{2}$ Departament de F\'{\i}sica, Universitat Polit\`{e}cnica de Catalunya, c/Esteve Terrades 5, 08860 Castelldefels, Spain\\
$^{3}$ Institut d'Estudis Espacials de Catalunya, Ed. Nexus-201, c/Gran Capit\`a 2-4, 08034 Barcelona, Spain\\
$^{4}$ Department of Applied Mathematics, University of Colorado, Boulder, CO 80309-0526, USA\\
}
\date{Accepted 2022 November 10. Received 2022 November 04; in original form 2022 September 16}

\pubyear{}

\begin{document}
\label{firstpage}
\pagerange{\pageref{firstpage}--\pageref{lastpage}}
\maketitle

\begin{abstract}
  The third data release of {\it Gaia} has provided low resolution spectra for $\sim$\,100\,000 white dwarfs (WDs) that, together with the excellent photometry and astrometry, represent an unrivalled benchmark for the study of this population. In this work, we first built a highly-complete volume-limited sample consisting in 12\,718 WDs within 100\,pc from the Sun. The use of VOSA tool allowed us to perform an automated fitting of their spectral energy distributions to different atmospheric models. In particular, the use of spectrally derived J-PAS photometry from {\it Gaia} spectra led to the classification of DA and non-DA WDs with an accuracy $>$\,90\%, tested in already spectroscopically labelled objects. The excellent performance achieved was extended to practically the whole population of WDs with effective temperatures above 5500\,K. Our results show that, while the A branch of the {\it Gaia} WD Hertzsprung-Russell diagram is practically populated by DA WDs, the B branch is largely formed by non-DAs (65\%). The remaining 35\% of DAs within the B branch implies a second peak at $\sim$\,0.8\,\msun\ in the DA-mass distribution. Additionally, the Q branch and its extension to lower temperatures can be observed for both DA and non-DA objects due to core crystallisation. Finally, we derived a detailed spectral evolution function, which confirms a slow increase of the fraction of non-DAs as the effective temperature decreases down to 10\,500\,K, where it reaches a maximum of 36\% and then decreases for lower temperatures down to $\sim$\,31\%.
\end{abstract}

\begin{keywords}
white dwarfs -- stars: evolution -- Galaxy: stellar content-- astronomical data bases: miscellaneous -- catalogues -- virtual observatory tools 
\end{keywords}


\section{Introduction}

The excellent quality of astrometric and photometric data provided by the ESA mission {\it Gaia} has been recently improved by the publication in its third data release (DR3) of nearly 200 million spectra \citep{DeAngeli22}. These spectra, although having a very low resolution (R\,$\approx$\,60), represent an invaluable source of information for a wide range of stellar objects in our Galaxy. In particular, nearly $100\,000$ spectra correspond to white dwarfs \citep{Montegriffo2022}.

As it is well known, white dwarfs are stellar remnants of low- to intermediate-mass main sequence stars \citep{Althaus2010}. Since nuclear fusion processes have ceased after the earlier stages of their lives, white dwarfs are stellar objects supported against gravity by Fermi electron pressure and they are doomed to a long-standing cooling process. As this physical process is relatively well understood and some white dwarfs can reach very old ages (10 Gyr or older), they can be used as reliable cosmochronometers \citep{Fontaine2001}, revealing valuable information about the history and evolution of our Galaxy \citep{GB2016}.

The source of energy in the deep interior of white dwarfs -- basically due to the gravothermal energy released by the ions and eventually provided by core crystallisation, phase separation, and sedimentation of minor species among other processes \citep[see][for a recent review]{Isern2022} -- is controlled by a thin partially degenerate layer through which the heat is radiated away. In the canonical model, this layer is formed by helium with a mass around $10^{-2}\,$\msun, representing less than 2\% of the total white dwarf mass, and in most of the cases ($\sim$\,80\%), an extra thinner layer of hydrogen with a mass between $10^{-15}$-$10^{-4}\,$\msun\ that lies on top of the helium one.

Observationally, characterisation of white dwarfs has been carried out mainly from spectroscopy, based on the atmospheric observed features, for instance through the identification of Balmer and helium lines, among others \citep{Sion1983}. If the white dwarf spectrum presents hydrogen lines the white dwarf is labelled as DA, while if the spectrum presents absorption helium lines, \ion{He}{I} or \ion{He}{II}, the white dwarf is named DB or DO, respectively. Although less common, it is possible to identify some metals within white dwarf atmospheres, such as carbon, named DQ, or other heavy elements such as \ion{Ca}{II} or \ion{Fe}{II}, assigned to the DZ spectral type. Some white dwarfs can also present featureless spectra, named in this case as DC. This last condition is usually satisfied by cool objects, in particular for hydrogen-dominated atmospheres with effective temperatures ($T_{\rm eff}$) below $\sim$\,5000\,K, or for helium-dominated atmospheres with $T_{\rm eff}$\,$\lappr$\,11\,000\,K. This initial list of spectral types has been extended to include mixed cases. Thus, we can find hydrogen-dominated atmospheres with some helium features, named DAB, or the contrary case, DBA, or any possible combination such as DBQ, DAZ, and so on. Also possible is the detection of a magnetic field or the existence of some variability in the white dwarf spectrum, adding in these cases and extra H or V, respectively, to the basic spectral type, e.g., DAH, DAV, DQV. All in all, we can summarize the spectral classification of the white dwarf population in two main classes: those with a hydrogen-dominated atmosphere (which is by far the most common case), DA, and the rest, named as non-DA. In any case, the importance of determining the atmospheric content of a white dwarf is crucial, not only for understanding the physical processes in their evolution, but also to derive reliable individual parameters such as masses, temperatures, and ages.

Spectral identification has historically relied on visual inspection of spectra. However, with the advent of large spectroscopic surveys, such as Sloan Digital Sky Survey \citep[SDSS;][]{York2000}, Large Sky Area Multi-Object Fibre Spectroscopic Telescope \citep[LAMOST;][]{Zhao2012}, {\it Gaia} \citep[][]{GaiaCollaboration16a} and many others, new automated data analysis techniques are required. In this sense, the Spanish Virtual Observatory (SVO) developed a powerful tool, the {\it Virtual Observatory Spectral energy distribution Analyzer} \citep[VOSA\footnote{\url{http://svo2.cab.inta-csic.es/theory/vosa}};][]{Bayo08}. VOSA is a Virtual Observatory (VO) tool that allows the user to build spectral energy distributions (SEDs) from both private and public photometric data within the VO, of thousands of objects at a time, and to derive their physical properties by comparing the observed SEDs to different collections of theoretical models. Moreover, it can estimate the goodness of the fitting of the model to the observed data.

In this work we aimed at thoroughly analyzing the 100\,pc nearly-complete volume-limited sample of white dwarfs with the main objective of individually classifying them based on their main spectral type class (DA or non-DA). The paper is, thus, organized as follows. After this introduction, in Section \ref{sample} we describe our methodology for building our volume-complete observed sample of white dwarfs within 100\,pc from the Sun. In Section \ref{s:spectral}, we detail the different white dwarf atmospheric models used, and how the SEDs were built from VO photometry or from synthetic photometry derived from the {\it Gaia} spectra. In Section \ref{s:class}, we introduce a set of white dwarf spectral type estimators based on the VOSA fitting and also, for comparative purposes, others estimators based on data found in the literature. In Section \ref{s:phys-prop} we apply the best estimator found on the basis of an already classified population of white dwarfs to the entire population, thus obtaining reliable distributions of stellar parameters, such as mass and effective temperature. Finally, we summarize our main results and conclusions in the last section.


\section{The {\it Gaia}-DR3 100\,pc white dwarf sample}
\label{sample}

In this section we describe our methodology for identifying white dwarf candidates within 100\,pc in the {\it Gaia}-DR3 catalogue.

In \cite{Jimenez-Esteban18} (hereafter JE18), we demonstrated that the completeness of the 100\,pc white dwarf sample selected using the second {\it Gaia} data release \citep[DR2;][]{Gaia2018} reached $\sim$\,94\% for a parallax relative error lower than $\sim$\,10\%. We also noted that smaller samples in volume are not necessarily more complete due to the Lutz-Kelker bias \citep{LutzKelker1973} and that for larger volumes, the completeness also decreases due to the magnitude limit of the {\it Gaia} satellite \citep{GaiaCollaboration16a}. As a consequence, we focused in this work on the construction of a white dwarf catalogue up to 100\,pc by selecting {\it Gaia}-DR3 sources with parallaxes larger than 10 mas, considering the errors, and imposing relative errors smaller than 10\% in parallax and in $BP$- and $RP$-band photometry. Photometric errors in the $BP$ and $RP$ bands were especially important for faint sources in crowded areas. Since the $G$-band photometry had lower error than the other two bands, we did not impose any cut in the $G$-band. Thus, we queried the {\it Gaia}-DR3 catalogue\footnote{\url{http://gea.esac.esa.int/archive/}} using the following criteria:
\begin{itemize}
  \item $\omega-3\sigma_{\omega}\ge10$\,mas and $\omega/\sigma_{\omega}\ge10$
\item $F_{\rm BP}/\sigma_{F_{\rm BP}}\ge10$ and $F_{\rm RP}/\sigma_{F_{\rm RP}}\ge10$
  \item RUWE<1.4
 \end{itemize}

The cut in the {\it RUWE} (Renormalised Unit Weight Error) parameter prevented against poor astrometric solutions \citep{Lindegren20a}. These criteria guaranteed that the sources were indeed within the 100\,pc volume. In addition, we estimated the corrected $BP$ and RP flux excess factor ($C^*$) and its scatter ($\sigma_{C^{*}}$), and applied a 3$\sigma_{C^{*}}$ cut following the recommendations by \cite{Riello20}. This criterion excluded some objects in the coolest end of the white dwarf sequence, where the oldest white dwarfs are expected. However, this guaranteed the consistency between the $G$-band and the $BP$ and RP photometry. 

\begin{figure}
  \includegraphics[width=\linewidth]{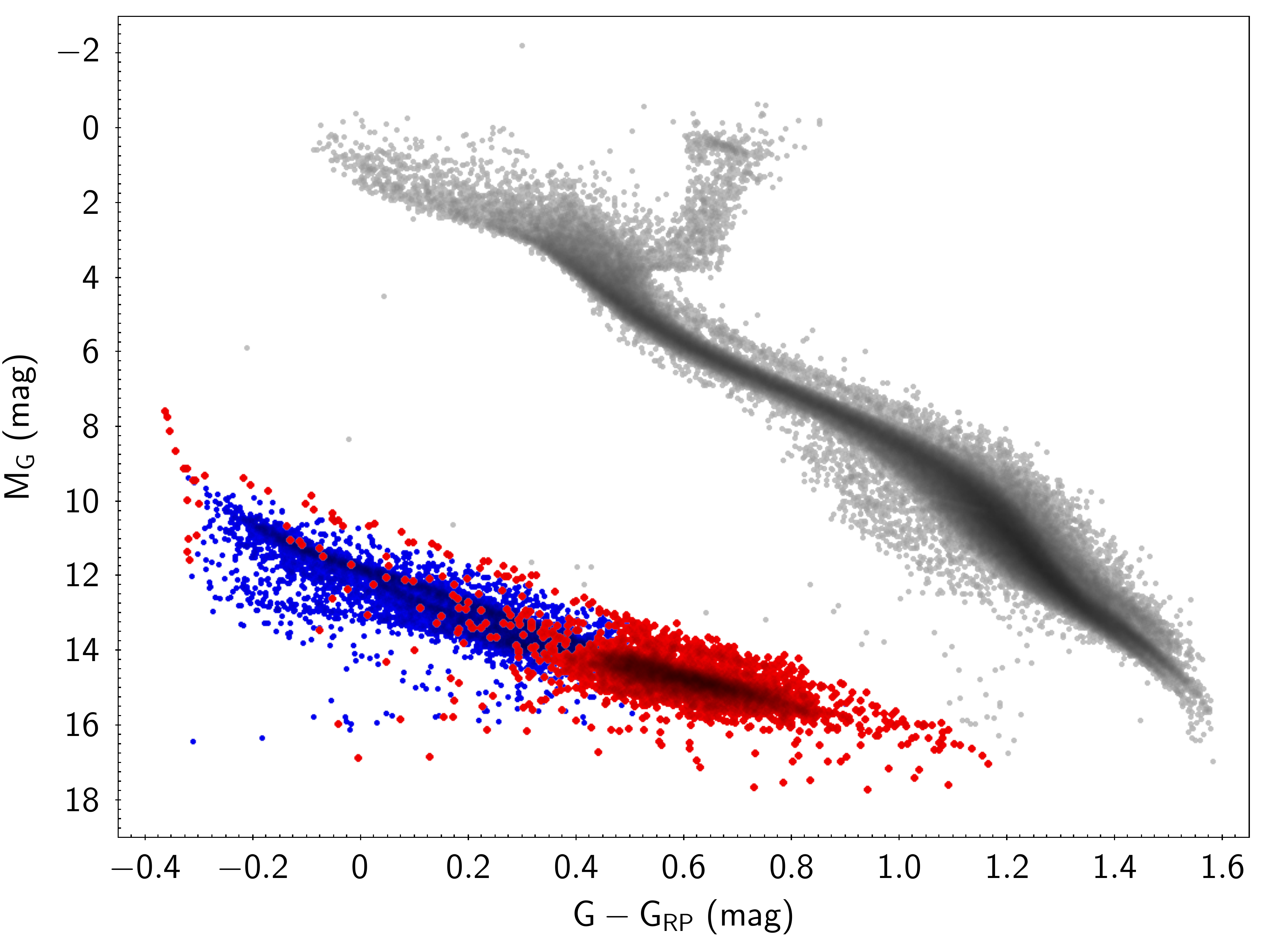}\par 
  \caption{{\it Gaia} Hertzsprung--Russell diagram of all {\it Gaia} sources within 100\,pc considered in this work. White dwarfs already identified in the {\it Gaia}-DR2 catalogue by \citet{Jimenez-Esteban18} are plotted in blue, and in red the new candidates identified in the {\it Gaia}-DR3 catalogue. Those stars fulfilling our 100\,pc selection criteria but not selected as white dwarf candidates are shown in grey.}
  \label{fig:HRD}
\end{figure}

We show in Fig.\,\ref{fig:HRD} the {\it Gaia} Hertzsprung--Russell (HR) diagram for all sources within 100\,pc with good photometric and astrometric data, as defined by the above criteria. Since the $BP$-band photometry has a known bias for red and faint sources \citep{Fabricius20}, we used the $G-G_\mathrm{RP}$ colour instead of the more commonly adopted $G_\mathrm{BP}-G_\mathrm{RP}$. The 10\% of uncertainty of the {\it Gaia} fluxes corresponds to an uncertainty of $\sim$\,0.14 mag in the {\it Gaia} colour. On the other hand, the 10\% uncertainty in the parallax corresponds to an uncertainty of $\sim$\,0.2 mag in the absolute $G$-band magnitude. This diagram clearly shows the {\it Main Sequence}, the {\it Red Giant Branch}, and the {\it White Dwarf Sequence}. Based on the location within the colour-magnitude diagram, we selected 12\,718 {\it Gaia}-DR3 sources located in the {\it White Dwarf Sequence} (blue and red dots in Fig.\,\ref{fig:HRD}), thus representing our 100\,pc sample. It is worth noting that, 7480 of these sources (blue) were already in the {\it Gaia}-DR2 white dwarf catalogue\footnote{\url{http://svo2.cab.inta-csic.es/vocats/v2/wdw/}} by JE18, and that 5238 (red) are new incorporations from the {\it Gaia}-DR3 catalogue.

The {\it Gaia}-DR2 white dwarfs are located in Fig.\,\ref{fig:HRD} at bluer colour ($G-G_\mathrm{RP}\la$0.5\,mag), while most of the newly identified have redder colours up to $G-G_{RP}\sim$1.2\,mag. This is because JE18 imposed a colour limit of $G_\mathrm{BP}-G_\mathrm{RP}<0.8$ mag, corresponding to an T$_\mathrm{eff}$ of $\sim$\,6000\,K, to avoid contamination by field stars with bad {\it Gaia}-DR2 astrometry. Thanks to the improvement in the astrometry measurements of the {\it Gaia}-DR3, this colour cut is no longer needed, allowing us to study the red part of the {\it White Dwarf Sequence}. There is also a number of hotter white dwarfs ($G_\mathrm{BP}-G_\mathrm{RP}<0.8$ mag) that were not included in the {\it Gaia}-DR2 white dwarf Catalogue by JE18. 51 of them are located in the HR diagram above the box defined in that work. The rest was not included probably due to lower quality of the {\it Gaia}-DR2 astrometry and/or photometry. On the other hand, there are 139 white dwarf candidates within 100\,pc identified by JE18 from the {\it Gaia}-DR2 Catalogue which are now excluded from the new catalogue using {\it Gaia}-DR3. This is probably due to a wrong astrometric solution in the {\it Gaia}-DR2 catalogue.

In conclusion, the catalogue of white dwarfs within 100\,pc presented in this work, not only extends to redder colours than the previous one, but also discards misidentified objects and includes blue missing sources. This catalogue, contining the basic information of the selected sources, including the spectral classification (Sect.\,\ref{s:class}) and the physical parameters (Sect.\,\ref{ss:stellar}) derived in this work, can be accesed electronicaly through our online catalogue service (see Appendix\,\ref{append}).

\subsection{Contamination} 

\begin{table}
\caption{Available SIMBAD classification for 12\,454 of our selected {\it Gaia}-DR3 white dwarfs. Classifications in agreement with the white dwarf nature of our selected candidates are listed above and shown in italics. The estimated contamination is $\lappr$\,0.1\%.}
\begin{center}
\begin{tabular}{cr}
\hline \hline
SIMBAD class & N (Candidates)\\
\hline
\it{white dwarf}   & 3640 (7242) \\
\it{Blue object} & 4 \\
\it{Star} & 80 \\
\it{High proper motion} & 1449 \\
\it{In Binary / Multiple system} &  18 \\
\hline
Hot subdwarfs     & 12 (2) \\
Cataclysmic Variable   & 4 (1) \\
Variable star &  1 \\
Emission-line star   & 1 \\
\hline \hline
\end{tabular}
\end{center}
\label{t:class}
\end{table} 


We used the SIMBAD database to evaluate the contamination in our white dwarf catalogue by other types of sources. We searched for SIMBAD counterparts within 3\arcsec\ of our {\it Gaia} white dwarf candidates using the {\it Gaia}-DR3 coordinates translated to epoch J2000 and corrected from {\it Gaia}-DR3 proper motions. Approximately 98\% (12\,454) of the sample had a counterpart in SIMBAD, all of them with a classification, at least tentative. The results are shown in Table\,\ref{t:class}. If we assume the SIMBAD classifications to be correct, including those flagged as candidates, and taken into account the position of these sources in the {\it Gaia} colour-magnitude diagram (see Fig.\,\ref{fig:HRD}), one can conclude that most ($>$99.8\%) 
of them have classifications compatible with white dwarf nature. 

The highest source of contamination seems to be hot subdwarfs ($\sim$\,0.1\%) and cataclysmic variables ($\sim$\,0.04\%).
However, although hot subdwarfs have similar colours than hot white dwarfs, they are more luminous and should occupy a different region in the HR diagram. So, the classification provided by SIMBAD is likely to be incorrect, and these objects are most probably misclassified white dwarfs. Thus, the expected contamination in our white dwarf catalogue is very low ($<$\,0.1\%).

\subsection{Completeness}

Our 100\,pc white dwarf sample is volume-limited and has a very low degree of contamination sample. However, it is subject to the astrometric and photometric constrains that were imposed to the {\it Gaia}-DR3 catalogue as explained above. Recently, from the {\it Gaia}-DR2 catalogue \cite{Hollands18} estimated a space-density of 4.49\,$\times$\,10$^{-3}$\,pc$^{-3}$ for the white dwarf population within 20\,pc from the Sun. Assuming the same space-density upto 100\,pc, we would expect $\sim$18.800 white dwarf in this volume. This implies that the completeness of our sample is $\sim$70\% of the entire 100\,pc white dwarf population. This estimation is in agreement with \cite{Torres22}.

Nevertheless, the level of completeness is not homogeneous along the {\it Gaia} colour range presented by the sample. Fig. 5 of \cite{Jimenez-Esteban18} shows the completeness as a function of the $G_\mathrm{BP}-G_\mathrm{RP}$ colour of a similar sample within 100\,pc obtained from the previous {\it Gaia}-DR2 catalogue. The completeness is almost 100\% for $G_\mathrm{BP}-G_\mathrm{RP}$\,$<$\,0\,mag and it continuously decreases to $\sim$70\% for the reddest colours $G_\mathrm{BP}-G_\mathrm{RP}$\,=\,0.8\,mag. It is expected that the completeness continues decreasing for redder colours than $G_\mathrm{BP}-G_\mathrm{RP}$\,=\,0.8\,mag. 

Our sample was built from the {\it Gaia}-DR3 catalogue, which has both better astrometry and photometry than the previous {\it Gaia}-DR2, so the completeness of the present sample is expected to be higher than the previous one. Consequently, we conclude that most of the missing sources are at the reddest colours, and the completeness of the spectroscopically classified sources ($G_\mathrm{BP}-G_\mathrm{RP}<0.86$ mag; see Sect.\ref{s:class}) is larger than 90\%.

Regarding the missing objects, one cause of missing sources is the confusion in the Galactic plane. Using our white dwarf sample, we compared the number of white dwarfs per square degree in the Galactic plane (|b|\,<\,10\,deg) and out of it. We found that the white dwarf sky density in the Galactic plane is 5\% lower than in the rest of the sky. Thus, we estimated $\sim$120 missing white dwarfs due to confusion. This is $\sim$1\% of our sample.

Another cause of missing white dwarfs is the number of double degenerate systems (DWD) which were not resolved by {\it Gaia} and were counted as single white dwarfs. \cite{Torres22} used different models to estimate the fraction of such systems. The best model estimated a fraction between 1\% and 3\% (depending on the common-envelope treatment) of unresolved DWDs within 100 pc. 

Finally, a third cause of missing sources are the binary systems made up by a white dwarf and a main sequence AFGK star, also known as Sirius-like systems. \cite{Holberg13} estimated that 8\% of the white dwarfs in the solar vecinity are members of such systems. Since many of them are now resolved by {\it Gaia}, we can used this value as an upper limit of the missing white dwarfs in our sample.

Summing up the contribution of missing sources, we conclude that less than 10\% of white dwarfs located within 100\,pc are not part of our catalogue in the colour range $G_\mathrm{BP}-G_\mathrm{RP}<0.86$ mag, while in the reddest end of the white dwarf sequence completeness decreases and the missing source would be mainly due to the limitations of {\it Gaia} capabilities in both astrometry and photometry.

\section{Spectral energy distribution analysis}
\label{s:spectral}

In this section we present the analyse of the SEDs of our 100\,pc white dwarf sample built from public photometric data and from {\it Gaia} spectra. Our purpose was to classify them in two main spectral classes: DA and non-DA. To that end, we used two different grids of white dwarf atmosphere models: hydrogen-dominated atmospheres (type DA) and helium-dominated atmospheres (note that even though DB models were used, others spectral types were also expected and we therefore named this group as type non-DA). We did not account for interstellar extinction since our sample is located at short distances.

\subsection{White dwarf atmospheric models}
\label{s:models}
The collection of models used in the analysis adopt LTE (local thermodynamic equilibrium), hydrostatic equilibrium, plane-parallel, one-dimensional structure, and convection with the mixing-length approximation where appropriate. The DA models consider pure hydrogen composition, and the non-DA models consider helium with a small trace of hydrogen (log N(H)/N(He) = -6). The version of the mixing length is ML2 \citep[][]{Tassoul1990}, with the ratio of mixing-length to pressure scale height 0.7 in the DA and 1.25 in the non-DA white dwarfs. Basic methods and data are described in \citet{Koester10}. More recently many improvements were implemented, the most important being: improved treatment of molecules, non-ideal effects, unified line broadening theories for the strong lines of MgI, MgII, CaI, CaII, and Lyman alpha. Also, the hydrogen Stark profiles by \citet{Tremblay2009} and Tremblay (2015, priv. comm) are included.

The DA grid contains 1260 model spectra and covers effective temperatures from 3000 to 20\,000\,K in steps of 250\,K (note that below $\sim$\,5000\,K these white dwarfs are DCs), from 20\,000 to 30\,000\,K in steps of 1000\,K, and from 30\,000 to 40\,000\,K in steps of 2\,000\,K. For each effective temperature, the surface gravities range from 6 to 9.25\,dex in steps of 0.25\,dex. The non-DA grid contains 666 model spectra and, in this case, the effective temperatures cover the 5500 to 20\,000\,K range in steps of 250\,K (below $\sim$\,10\,000\,K these white dwarfs are also featureless DCs), the 20\,000 to 30\,000\,K range in steps of 1000\,K, and the range from 30\,000\ to 40\,000\,K in steps of 2\,000\,K. For this grid, the surface gravities range from 7 to 9\,dex in steps of 0.25\,dex for each effective temperature value.

\subsection{SED from VO archives}
\label{ss:SEDVO}

To build the SEDs of the 100\,pc white dwarf sample from public archives within the VO, we collected data from different photometric catalogues, which are listed in Table\,\ref{t:surveys}. Detailed information of the adopted filters can be found using the SVO Filter Profile Service\footnote{Available at \url{http://svo2.cab.inta-csic.es/theory/fps/}.} \citep[][]{Rodrigo2012,Rodrigo2020}. In order to avoid misidentifications, the {\it Gaia} coordinates for all sources were corrected from proper motion and transformed to an epoch closer to the observation epoch for each of the queried survey. A 3\arcsec\ search radius was adopted. This allowed us to build the observational SEDs from the UV to the mid-IR wavelength range for most of the sources.

\begin{table*}
\caption{Photometry data from VO archives used to build the SEDs. The detailed information on each photometric filter can be obtained from the SVO Filter Profile Service \citep[][]{Rodrigo2012,Rodrigo2020}.} 
\begin{tabular}{lcccc}
\hline \hline
Survey & Facility/Instrument & Spectral range & Filters & Reference \\
\hline
GALEX UV GR6+7 & GALEX & UV & FUN,\,NUV & 1 \\
XMM-SUSS4.1 & XMM/OM & UV,\,visible & UVW2,\,UVM2,\,UVW1,\,U,\,B,\,V & 2 \\
{\it Gaia} DR3 & Gaia & visible & G$_{\rm BP}$,\,G,\,G$_{\rm RP}$ & 3 \\
APASS DR9 & --- & visible & B,\,v,\,g,\,r,\,i & 4 \\
J-PLUS DR1 & OAJ/T80Cam & visible & uJAVA,\,J0378,\,J0395,\,J0410,\,J0430,\,gSDSS & 5 \\
 & & & J0515,\,rSDSS,\,J0660,\,iSDSS,\,J0861,\,zSDSS & \\
Tycho-2 & TYCHO & visible & B, V & 6 \\
SDSS DR12 & SLOAN & visible & u,\,g,\,r,\,i,\,z & 7 \\
IPHAS DR2 & INT/WFC & visible & gR,\,Ha,\,gI & 8 \\
VPHAS$+$ DR2 & Paranal/OmegaCAM & visible & u,\,g,\,H$\alpha$,\,i,\,z & 9 \\
Pan-STARRS1 DR2 & PAN-STARRS & visible,\,near-IR & g,\,r,\,i,\,z,\,y & 10 \\
Dark Energy Survey DR1 & CTIO/DECam & visible,\,near-IR & g,\,r,\,i,\,z,\,Y & 11 \\
DENIS & -- & visible,\,near-IR & I,\,J,\,K$_{\rm s}$ & 12 \\
VISTA & Paranal/VIRCAM & visible,\,near-IR & Z,\,Y,\,J,\,H,\,K$_{\rm s}$ & 13 \\
UKIDSS & UKIRT/WFCAM & visible,\,near-IR & Z,\,Y,\,J,\,H,\,K & 14 \\
2MASS & 2MASS & near-IR & J,\,H,\,K$_{\rm s}$ & 15 \\
GLIMPSE I+II+3D & Spitzer/IRAC & mid-IR & I1,\,I2,\,I3,\,I4 & 16 \\
AllWISE & WISE & mid-IR & W1,\,W2,\,W3,\,W4 & 17 \\

\hline \hline
\end{tabular}
\begin{tablenotes}
\item References: [1] Revised catalog of GALEX UV sources \citep[GALEX UV GR6+7;][]{Bianchi17}; [2] XMM-Newton Serendipitous Ultraviolet Source Survey catalogue \citep[XMM-SUSS4.1;][]{Page12}; [3] {\it Gaia} DR3 \citep{GaiaCollaboration22}; [4] AAVSO Photometric All-Sky Survey \citep[APASS DR9;][]{Henden15}; [5] Javalambre Photometric Local Universe Survey \citep[J-PLUS DR1;][]{Cenarro19}; [6] Tycho-2 Catalogue \citep{Hog00a}; [7] Sloan Digital Sky Survey Photometric Catalogue \citep[SDSS DR12][]{Alam15}; [8] INT/WFC Photometric H-Alpha Survey of the Northern Galactic Plane Catalogue \citep[IPHAS DR2][]{Barentsen14}; [9] VST Photometric Halpha Survey of the Southern Galactic Plane and Bulge \citep[VPHAS+ DR2;][]{Drew16}; 
[10] Panoramic Survey Telescope and Rapid Response System \citep[Pan-STARRS1 DR2;][]{Chambers16}; [11] Dark Energy Survey DR1 \citep[Dark Energy Survey DR1;][]{DESC16}; [12] Deep Near Infrared Survey of the Southern Sky third release data \citep[DENIS][]{Denis05}; [13] Visible and Infrared Survey Telescope for Astronomy \citep[VISTA][]{Cross12}; [14] UKIRT Infrared Deep Sky Survey \citep[UKIDSS][]{Hewett06}; [15] 2MASS All-Sky Catalog of Point Sources \citep[2MASS;][]{Skrutskie06}; [16] Galactic Legacy Infrared Midplane Survey Extraordinaire Source Catalog \citep[GLIMPSE I+II+3D]{Spitzer2009}; and [17] Wide-field Infrared Survey Explorer AllWISE Data Release \citep[AllWISE][]{Cutri14}.
\end{tablenotes}
\label{t:surveys}
\end{table*}

For 463 white dwarf candidates, no additional photometry to {\it Gaia} data were found, so their SEDs could not be fitted with VOSA. For the other 12\,255 sources, each individual SED was fitted to both DA and non-DA white dwarf synthetic spectra outlined in Sect.\,\ref{s:models}. Of them, we obtained a reliable fit, defined as Vgf$_b$\,$<$\,15\footnote{Vgf$_b$: Modified reduced $\chi^{2}$, calculated by forcing $\sigma(F_\mathrm{obs})$ to be larger than $0.1\times F_\mathrm{obs}$, where $\sigma(F_\mathrm{obs})$ is the error in the observed flux ($F_{obs}$). This can be useful if the photometric errors of any of the catalogues used to build the SED are underestimated. Vgf$_b$ smaller than 10\,-\,15 is often perceived as a good fit.}, for 10\,447 sources when using either DA or non-DA model spectra. 1104 white dwarfs obtained good fits only when using DA model spectra, while 65 white dwarfs obtained good fits only when using non-DA model spectra. In total, we obtained a good fit for 11\,616 sources ($\sim$\,95\% of the fitted sample).

\begin{figure*}
  \includegraphics[width=0.98\columnwidth]{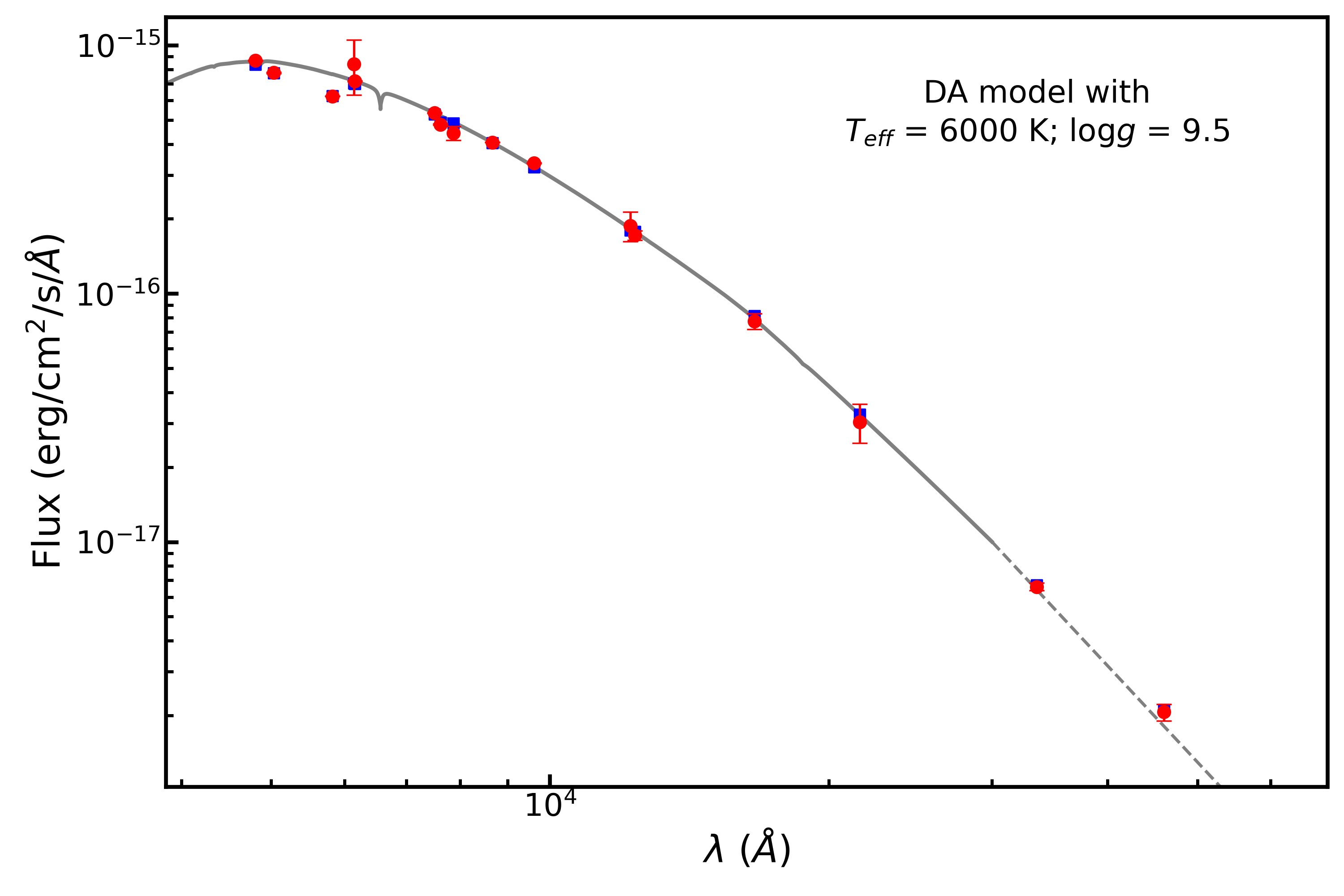}
  \includegraphics[width=0.98\columnwidth]{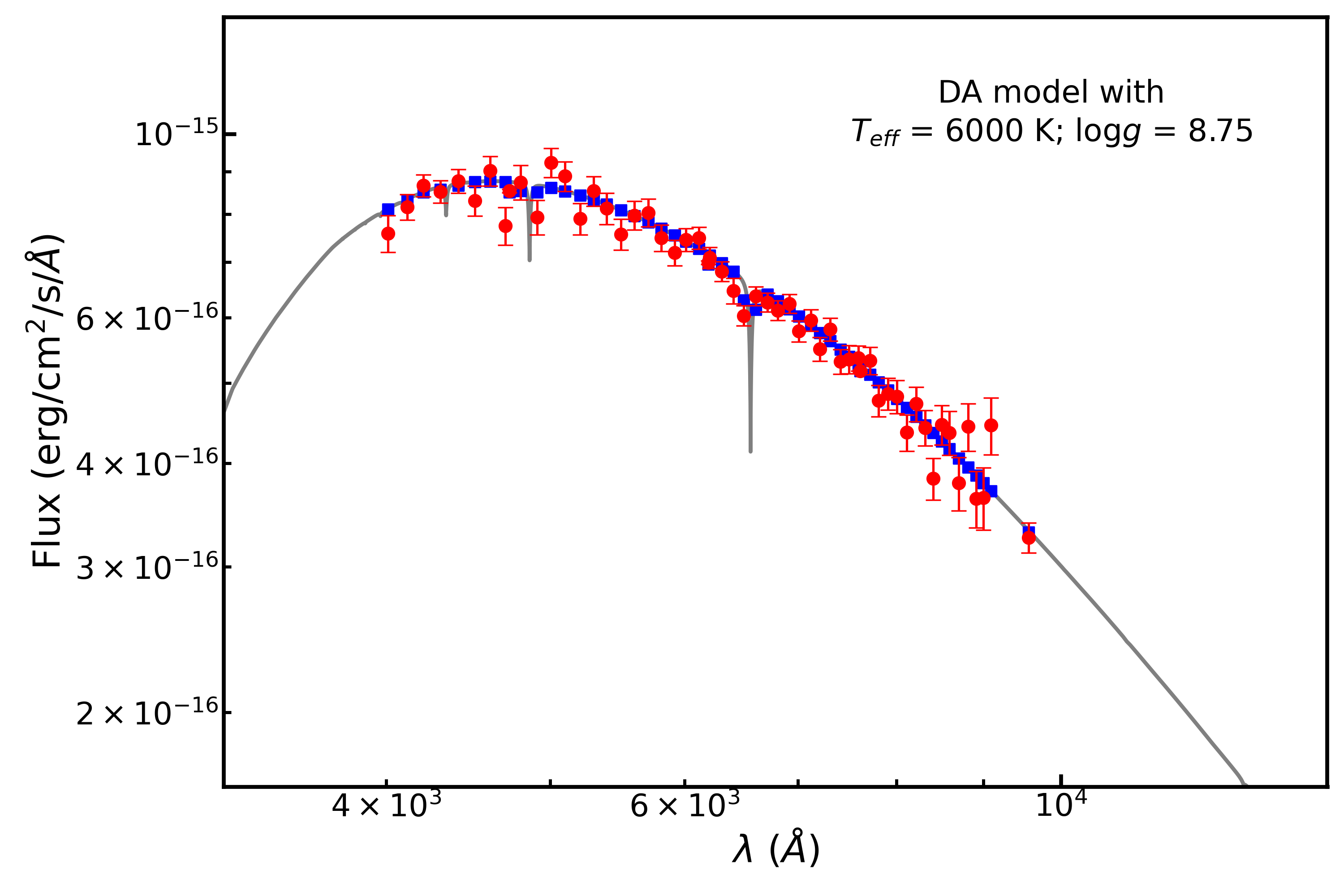} 
  \includegraphics[width=0.98\columnwidth]{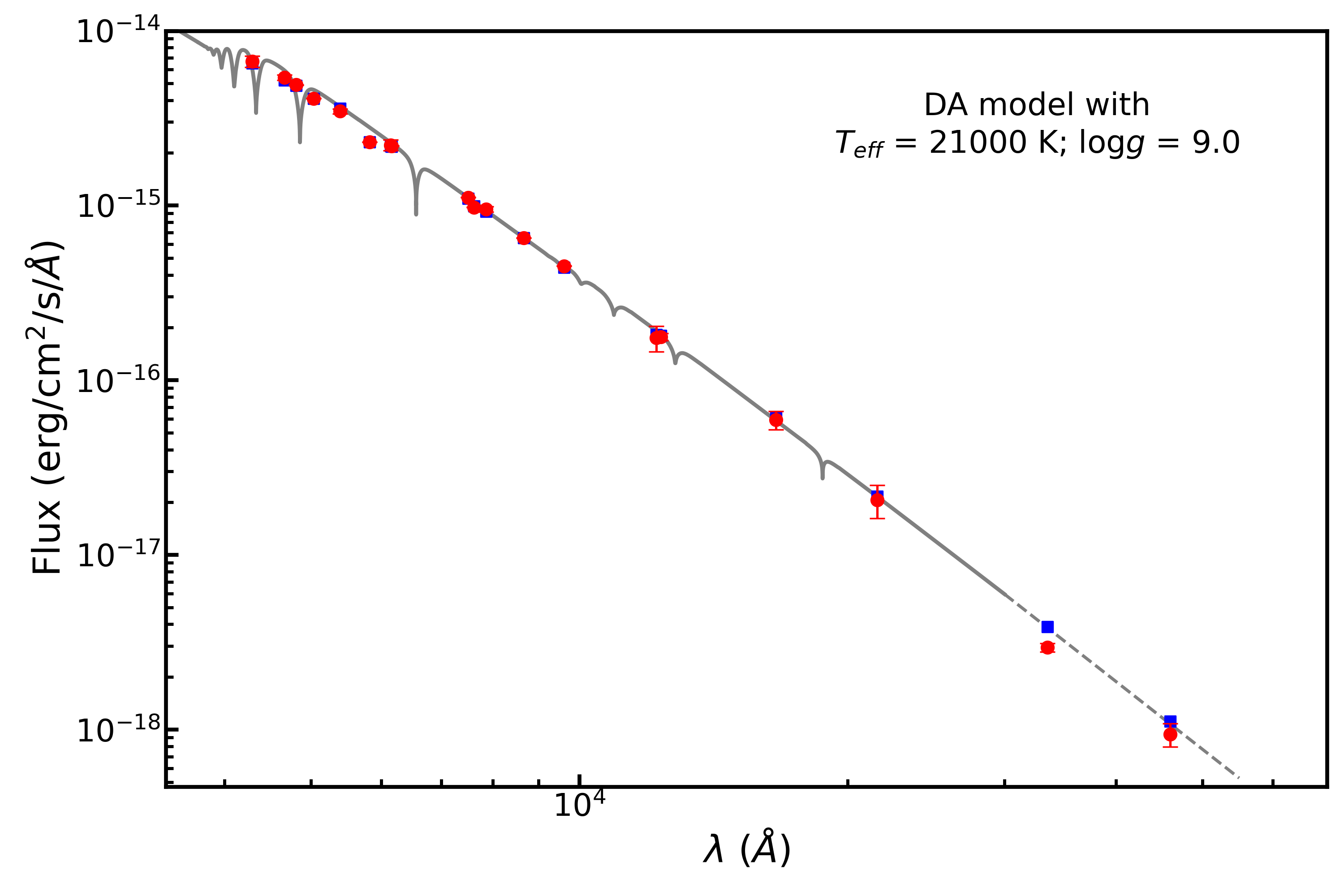} 
  \includegraphics[width=0.98\columnwidth]{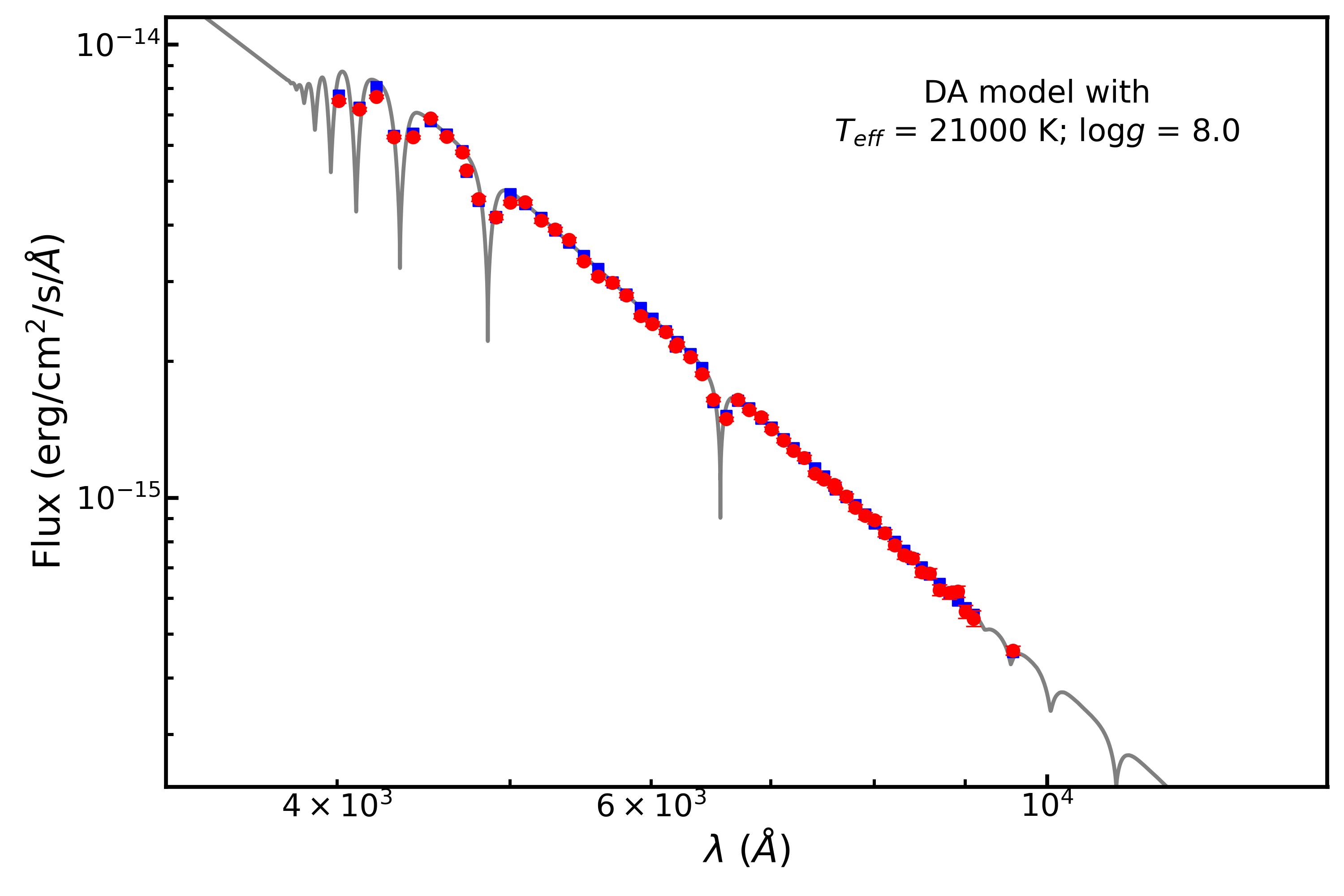}
  \includegraphics[width=0.98\columnwidth]{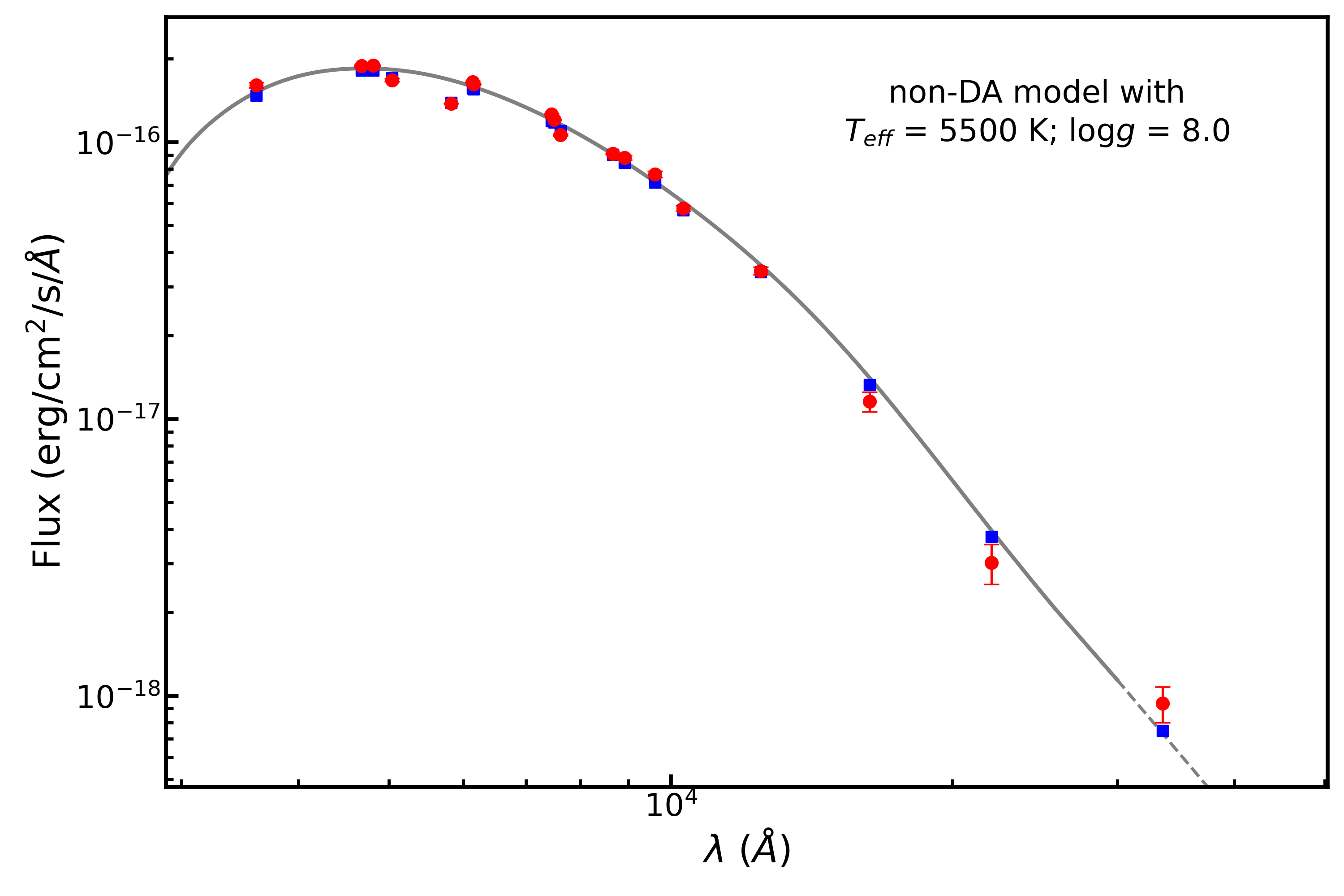}
  \includegraphics[width=0.98\columnwidth]{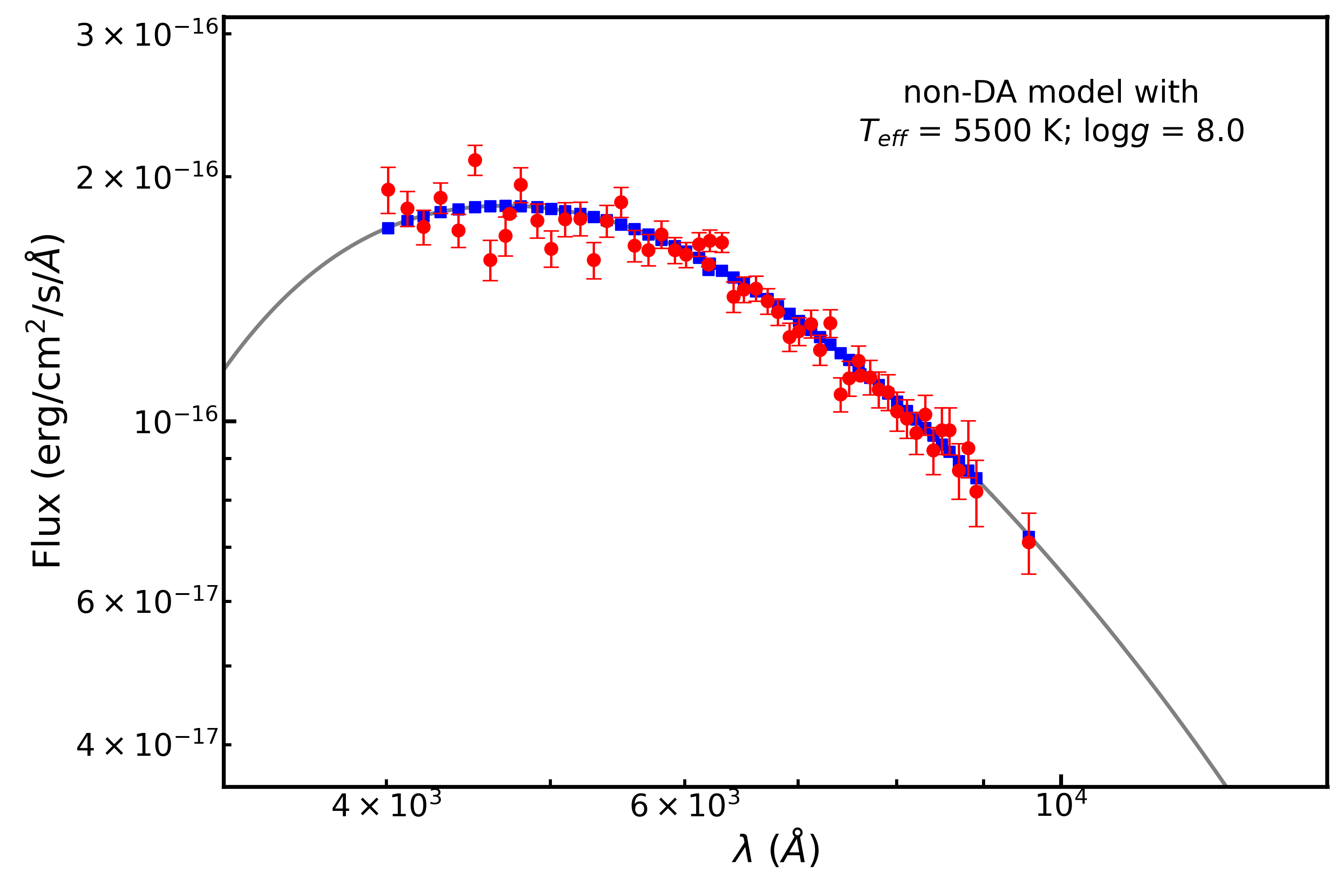}
  \includegraphics[width=0.98\columnwidth]{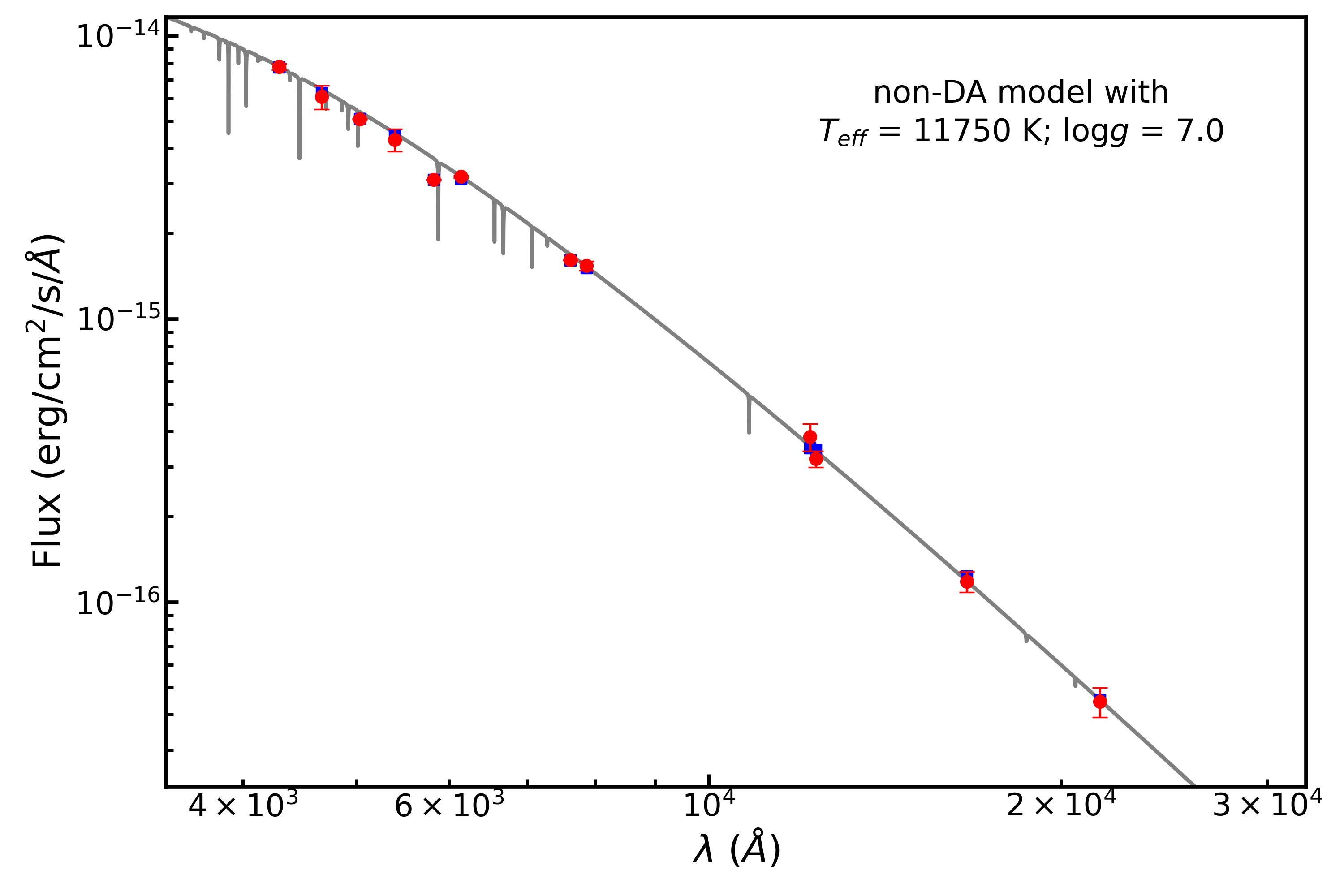}
  \includegraphics[width=0.98\columnwidth]{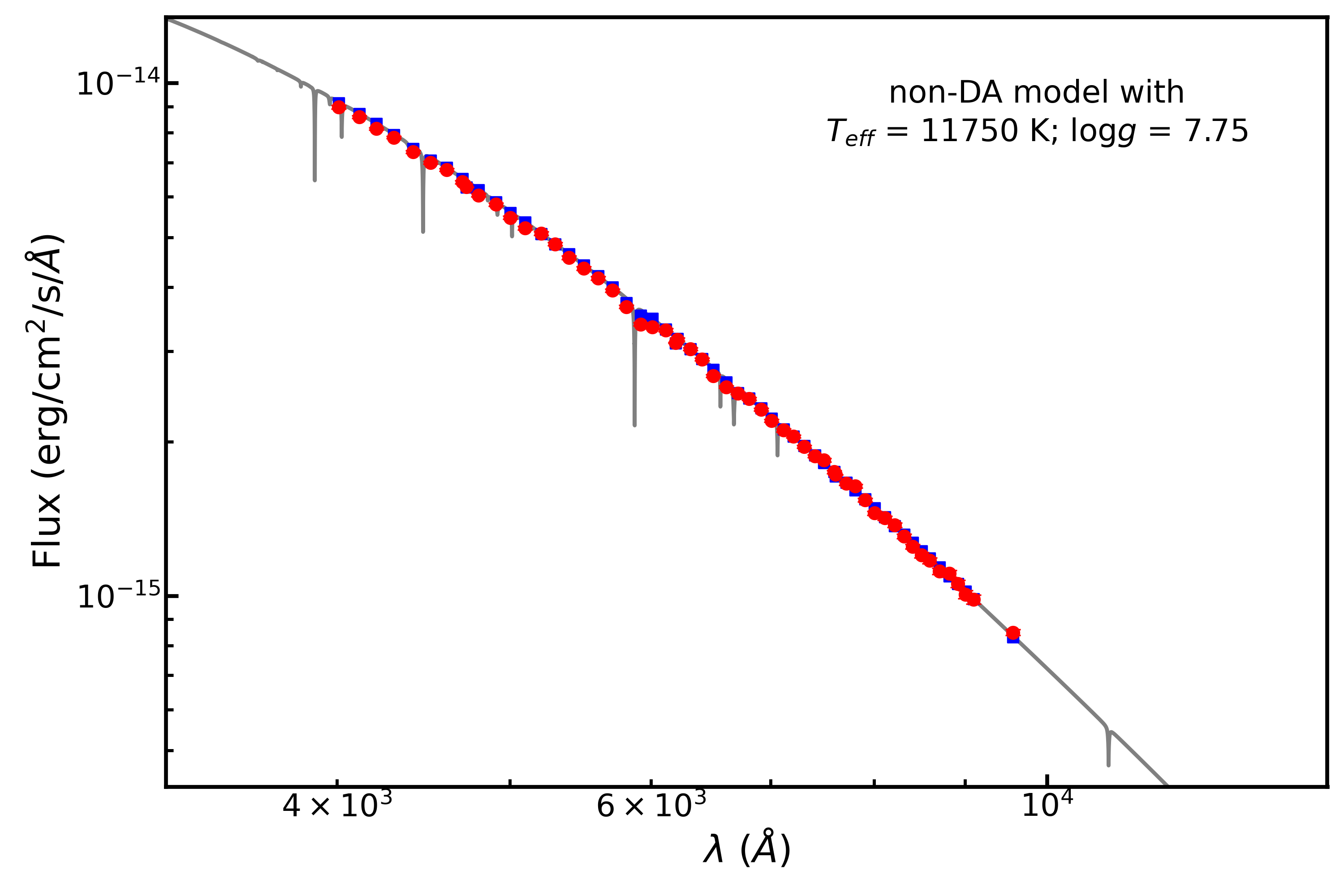}
\caption{Spectral energy distributions of four distinct white dwarfs, each row presents an object (from the top to the bottom: {\it Gaia}-DR3 3750749378584132992, {\it Gaia}-DR3 3020668542435696512, {\it Gaia}-DR3 2780434524599787136, and {\it Gaia}-DR3 5869567658943170048), illustrating DA and non-DA models with different effective temperatures and surface gravities. The red filled circles show the photometry available in VO archives (left panels) and the J-PAS photometry built from {\it Gaia} spectra considering all coefficients (right panels). The best-fit models are presented as grey solid line, where the blue squares are the synthetic photometry (calculated from the model) in each filter. The dashed line represents the Rayleigh-Jeans law adopted for longer wavelengths, when they are not covered by the models.}
  \label{f-SEDS}
\end{figure*}

For illustrative purposes, in the left panels of Fig.\,\ref{f-SEDS} we show the observational SEDs built from VO photometric catalogues (red circles), together with the synthetic photometry (blue squares) and the synthetic spectra (grey solid line) that best fit the observations. The two upper rows correspond to two white dwarfs which best-fit model is a DA, and the two bottom rows to two white dwarfs which best-fit model is a non-DA, with different effective temperatures and surface gravities. In the right panels of Fig.\,\ref{f-SEDS} we also show for the same sources the SEDs built with the J-PAS photometry obtained from {\it Gaia} spectra (see Sect.\,\ref{ss:SEDGaia}).

\subsection{SED from {\it Gaia} spectra}
\label{ss:SEDGaia}

The new {\it Gaia}-DR3 has provided low-resolution spectra for more than 200 millions sources \citep{DeAngeli22}. The spectra cover the optical to near-infrared wavelength range, from 330 to 1050\,nm, approximately. 12\,342 (97\% of the whole ample) white dwarfs in our sample had {\it Gaia} low-resolution spectra available. We used the Python package {\it GaiaXPy}\footnote{\url{https://www.cosmos.esa.int/web/gaia/gaiaxpy}} to construct the photometric SEDs of these sources from {\it Gaia} spectra using the {\it Javalambre-Physics of the Accelerating Universe Astrophysical Survey} (J-PAS; \citealt{Benitez14}) filter system \citep{Marin-Franch12}. This system is composed of 54 overlapping narrow-band (FWHM\,$\approx$\,145\,\AA) filters covering from 3780 to 9100\,\AA, plus 2 broad-band filters at the blue and red end of the optical range, and 4 complementary SDSS-like filters. In total, the 60 J-PAS filters provide a low-resolution ($R\approx$\,60) spectrum \citep{Bonoli21} similar to {\it Gaia}. However, due to the low signal-to-noise ratio of the {\it Gaia} spectra at bluer wavelengths \citep{Montegriffo2022}, we did not use the 4 J-PAS filters with effective wavelength shorter than 4000\,\AA, which reduces to 56 the number of J-PAS filters available for building the SEDs. Detailed information on the J-PAS filter system can be found at the SVO Filter Profile Service\footnote{\url{http://svo2.cab.inta-csic.es/theory/fps/index.php?mode=browse&gname=OAJ&gname2=JPAS}}. 

\begin{figure}
  \includegraphics[width=0.98\columnwidth]{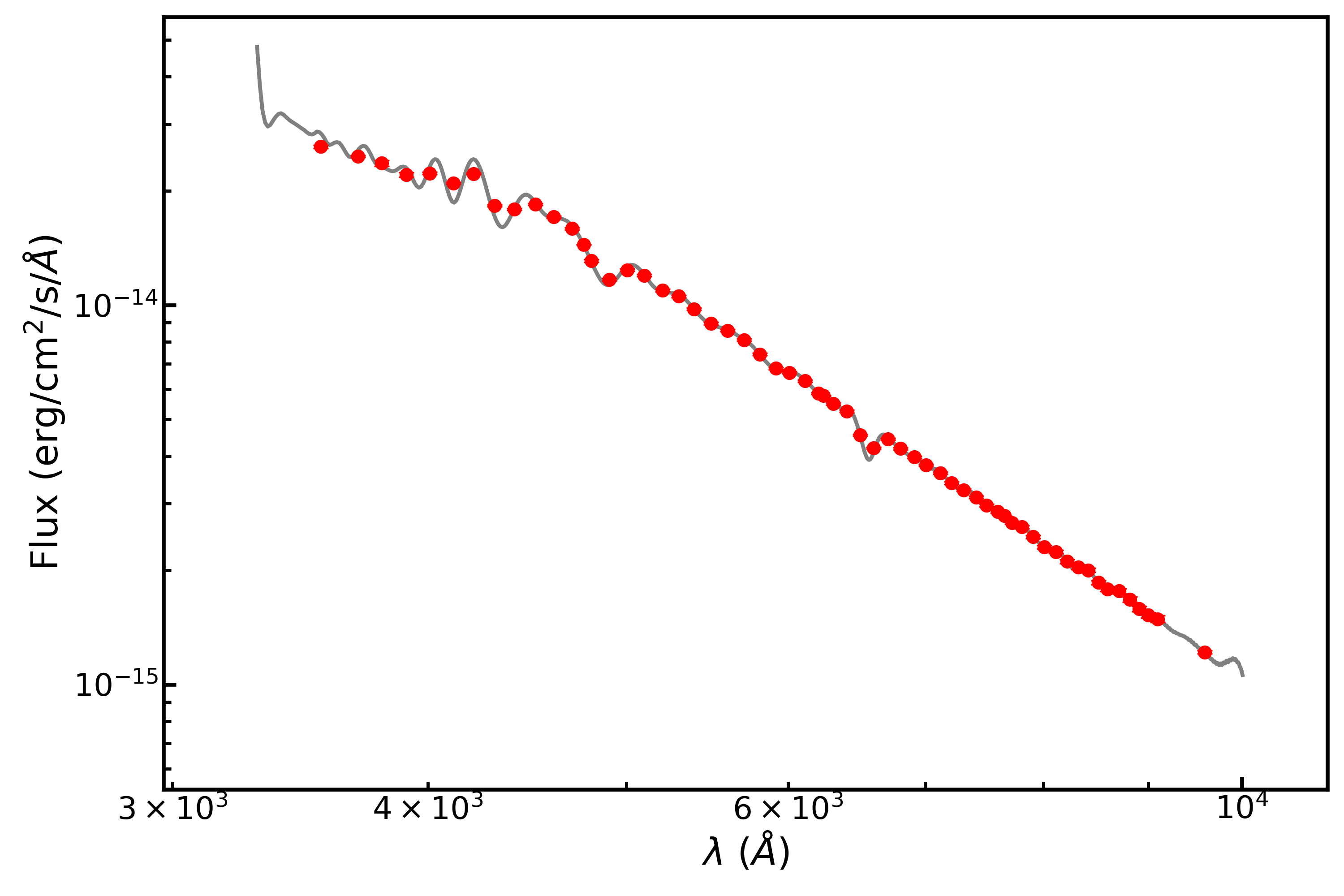} 
  \includegraphics[width=0.98\columnwidth]{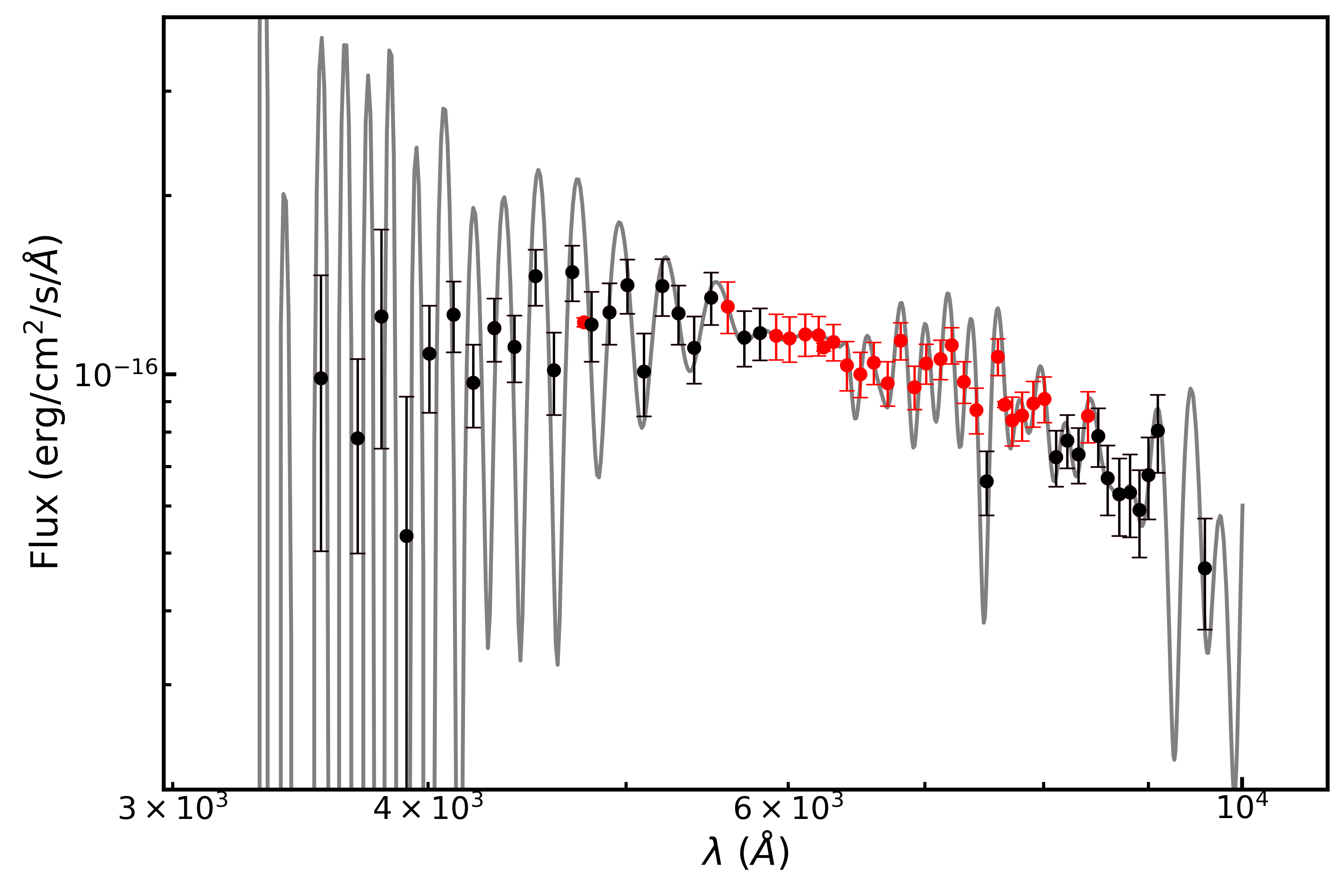}
\caption{Spectral energy distributions of two white dwarfs. {\it Gaia}-DR3 3683519503881169920 (upper panel) has one of the highest signal-to-noise ratio and blue colour ($G_\mathrm{BP}-G_\mathrm{RP}$\,$\sim$\,--0.32 mag). On the contrary, {\it Gaia}-DR3 5174110233491949312 (lower panel) has one of the lowest signal-to-noise ratio and red colour ($G_\mathrm{BP}-G_\mathrm{RP}$\,$\sim$\,0.86 mag). The filled circles show the J-PAS photometry built from {\it Gaia} spectra considering all coefficients. In red, those photometric points with a relative flux error lower than 10\% are shown, and then considered in the model fit, and in black are shown those not considered due to larger errors. The {\it Gaia} spectra are presented as grey solid lines.}
  \label{f-GJP}
\end{figure}

For each source in our sample with a {\it Gaia} spectrum, we built two SEDs using the J-PAS photometric system, one taking into account all the coefficients of the spectrum (GJP) and one taking into account only the relevant coefficients as provided by {\it Gaia}-DR3 (GJP-trunc) \citep{DeAngeli22}. We imposed a threshold in the photometric error of 10\% to each individual photometric measurement obtained with {\it GaiaXPy}. Thus, although most of the SEDs have 56 photometric points, in some noisy spectra the number of points is lower. For illustrative purposes, in Fig.\,\ref{f-GJP} we show two SEDs built considering all coefficients for a source with one of the highest and another with one of the lowest signal-to-noise ratio.

Of the 12\,342 {\it Gaia} spectra, we obtained 11\,447 GJP SEDs with more than 4 photometric points, the minimum number necessary to be fitted by VOSA, and 12\,324 GJP-trunc SEDs. Of them, we obtained a reliable fit (Vgf$_b$\,$<$\,15) for 11\,153 GJP and 11\,639 GJP-trunc SEDs, when using either DA or non-DA model spectra. 290 GJP and 678 GJP-trunc SEDs obtained good fits only when using DA model spectra, while no SED obtained a good fit only when using non-DA model spectra. In total, we obtained a good fit for 11\,443 GJP and 12\,317 GJP-trunc SEDs, almost all the analyzed SEDs obtained from {\it Gaia} spectra. To illustrate, we show some examples of the obtained GJP SEDs and their best fitting model in Fig.\,\ref{f-SEDS} (right panels).


\section{White dwarf Spectral classification}
\label{s:class}

In this section a set of white dwarf spectral class estimators is built. Most of these estimators were based on the result of the fitting of the SEDs with certain white dwarf atmospheric models as described in the previous Section. The performance of these estimators was validated through the spectroscopically confirmed white dwarfs from the Montreal White Dwarf Database \citep[MWDD;][]{Dufour2017}. 

\subsection{Spectral estimators}

Following the VOSA procedure, in which each SED was fitted to two models (a DA and a non-DA model), and for which VOSA derived two reduced $\chi^2$-values ($\chi^2_{\rm DA}$ and $\chi^2_{\rm non-DA}$, respectively), we built an estimator, $i$, that measures the probability of being a hydrogen-dominated atmosphere (DA) white dwarf, $P_{\rm DA}^{i}$, as 

\begin{equation}
\label{e:probDA}
P_{\rm DA}^{i}=\frac{1}{2}\left(\frac{\chi^2_{\rm non-DA}-\chi^2_{\rm DA}}{\chi^2_{\rm non-DA}+\chi^2_{\rm DA}}+1 \right)
\end{equation}

\noindent

Hence, we classified an object as a DA white dwarf if $P_{\rm DA}^{i}\ge \eta$, where $\eta$ is a threshold value, otherwise we considered it as a non-DA. In our analysis we adopted an initial value of $\eta=0.5$.

This estimator can be generalized by using any other $\chi^2$-values available in the literature. Hence, up to 6 estimators were analyzed in this work. A description of them is presented as follows, while a summary is listed in Table \ref{t:estimators}.

\begin{table*}
\caption{Summary of the white dwarf spectral estimator built for the present study.}
 \begin{tabular}{| l | c | c | c }
\hline \hline
Estimator name & Reference & Atmosphere models & mean number points per SED \\
 \hline
VOSA      & This work & Pure hydrogen \citep{Koester10} & 20 \\
      & & Nearly pure helium H/He= -6 \citep{Koester10} & \\
VOSA-GJP    & This work & idem & 56\\
VOSA-GJP-trunc & This work & idem & 56\\
GF21-I     & \cite{GentileFusillo21} & Pure hydrogen \citep{Tremblay2011,Kowalski10} & 3 \\
    & & Pure helium \citep{Bergeron2011} &  \\
GF21-II     & \cite{GentileFusillo21} & Pure hydrogen \citep{Tremblay2011,Kowalski10} & 3 \\
    & & Mixed H/He=-5 \citep{Tremblay2014,McCleery2020} &  \\
MON22      & \cite{Montegriffo2022} & * & 60 \\
\hline \hline
 \end{tabular}
\label{t:estimators}
\end{table*}

\subsubsection{The VOSA, VOSA-GJP, and VOSA-GJP-trunc estimators}

In Section\,\ref{ss:SEDVO} we described the use of VOSA for analyzing white dwarf SEDs. For those objects with a reliable fit, i.e., Vgfb<15, using both DA and non-DA atmospheric model, and by means of eq.\,\ref{e:probDA}, we built three different estimators. The first one, called VOSA, corresponds to the case that the SEDs were built from public archives within the VO (see Section \ref{ss:SEDVO}). In the case that the SEDs were derived from {\it Gaia} spectra (see Section \ref{ss:SEDGaia}), we defined two extra estimators: if all the coefficients of the white dwarf spectra were taken into account, our estimator is named VOSA-GJP; and in the case that the truncated version of the coefficients was adopted, our estimators is called VOSA-GJP-trunc.

\begin{figure*}
\centering
  \includegraphics[width=0.32\textwidth,trim=2 10 20 20, clip]{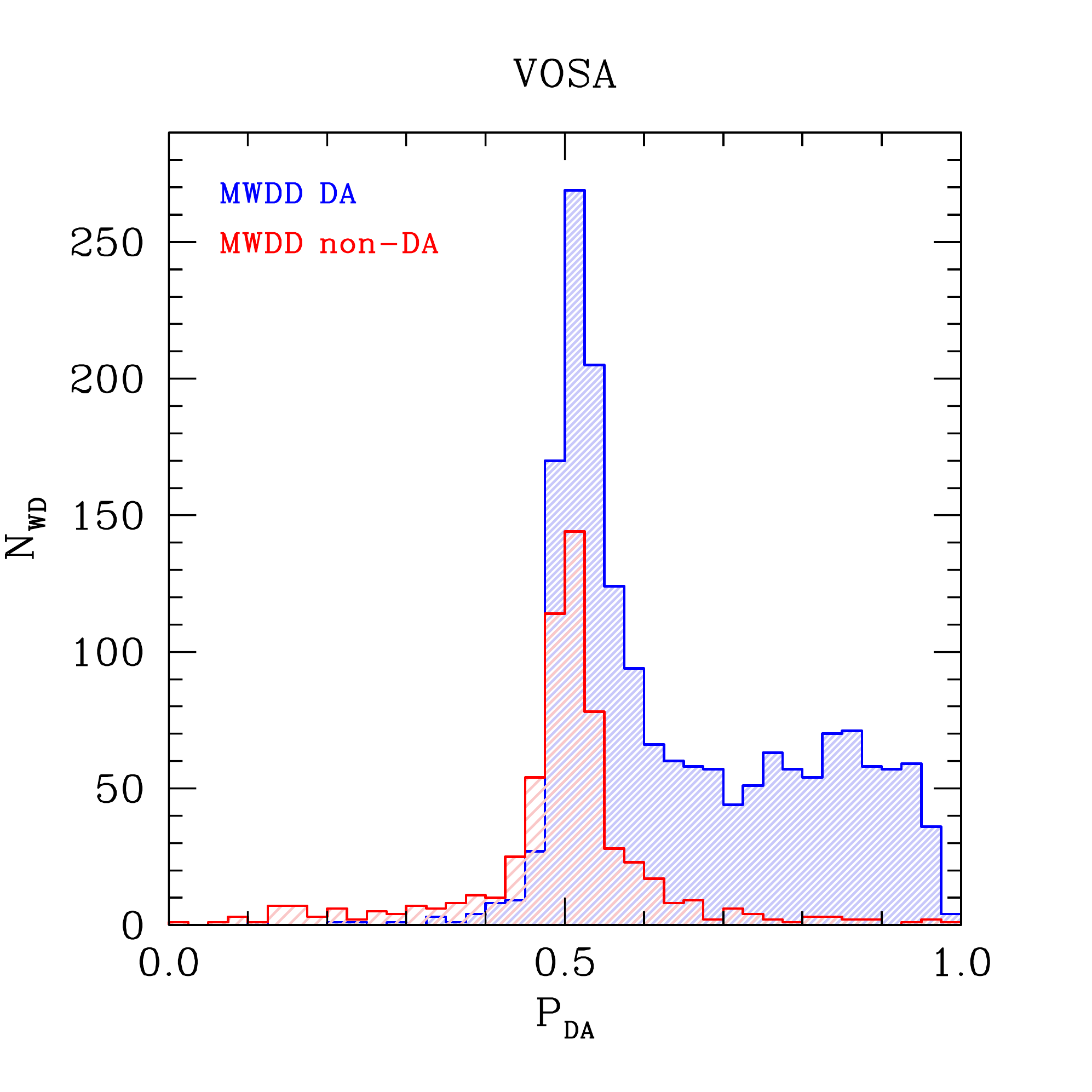}
  \includegraphics[width=0.32\textwidth,trim=2 10 20 20, clip]{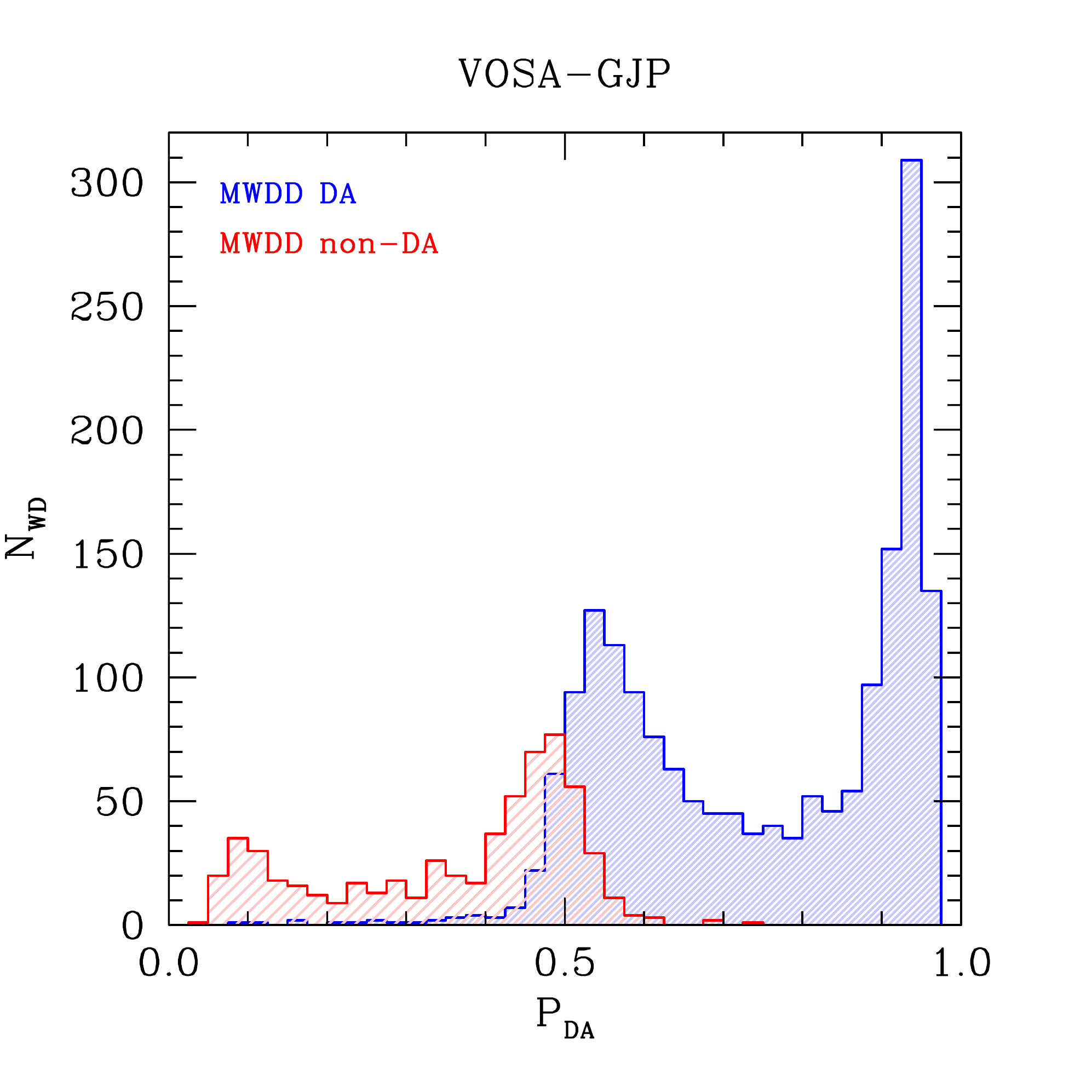}
  \includegraphics[width=0.32\textwidth,trim=2 10 20 20, clip]{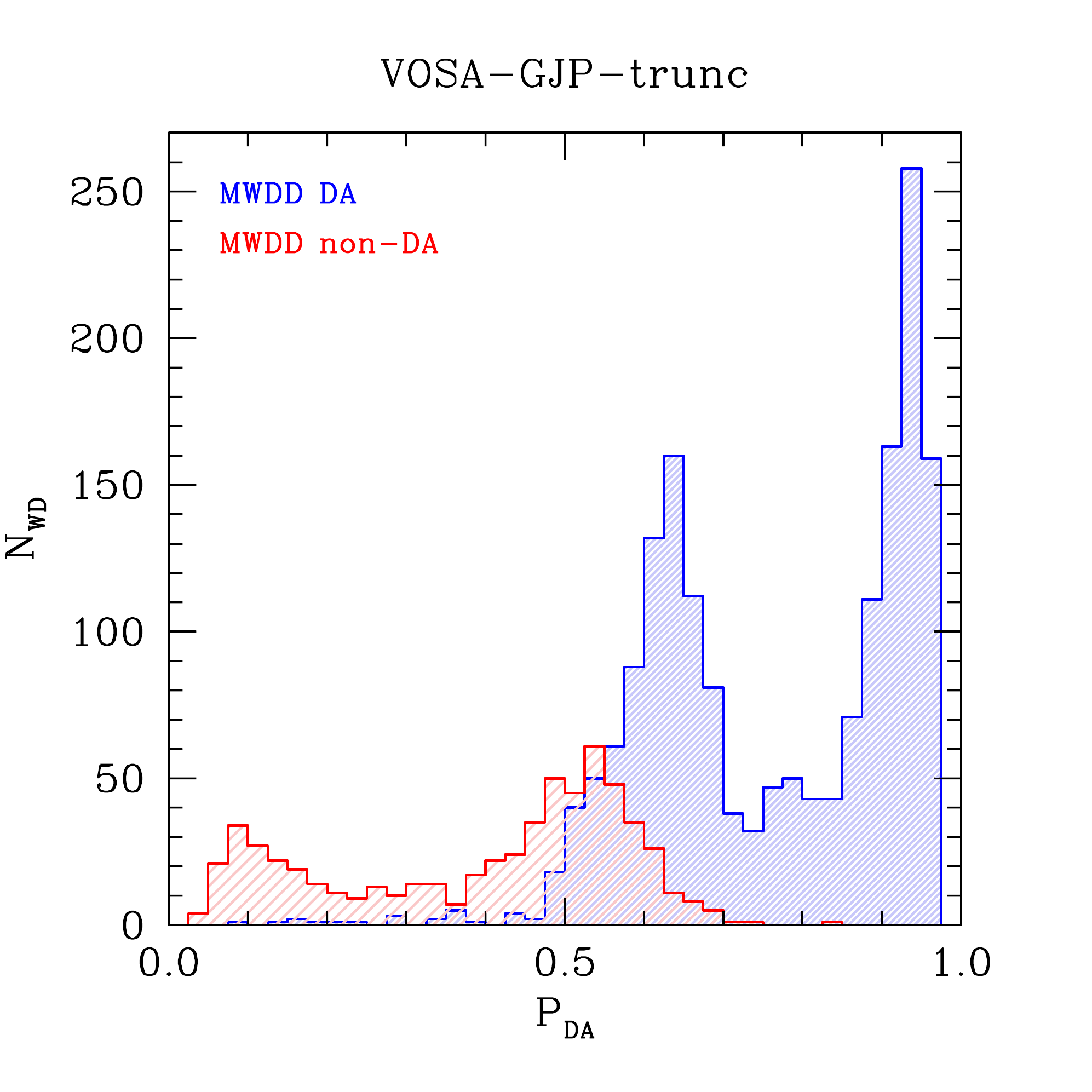}
       \vskip 0.4cm
  \includegraphics[width=0.32\textwidth,trim=2 10 20 20, clip]{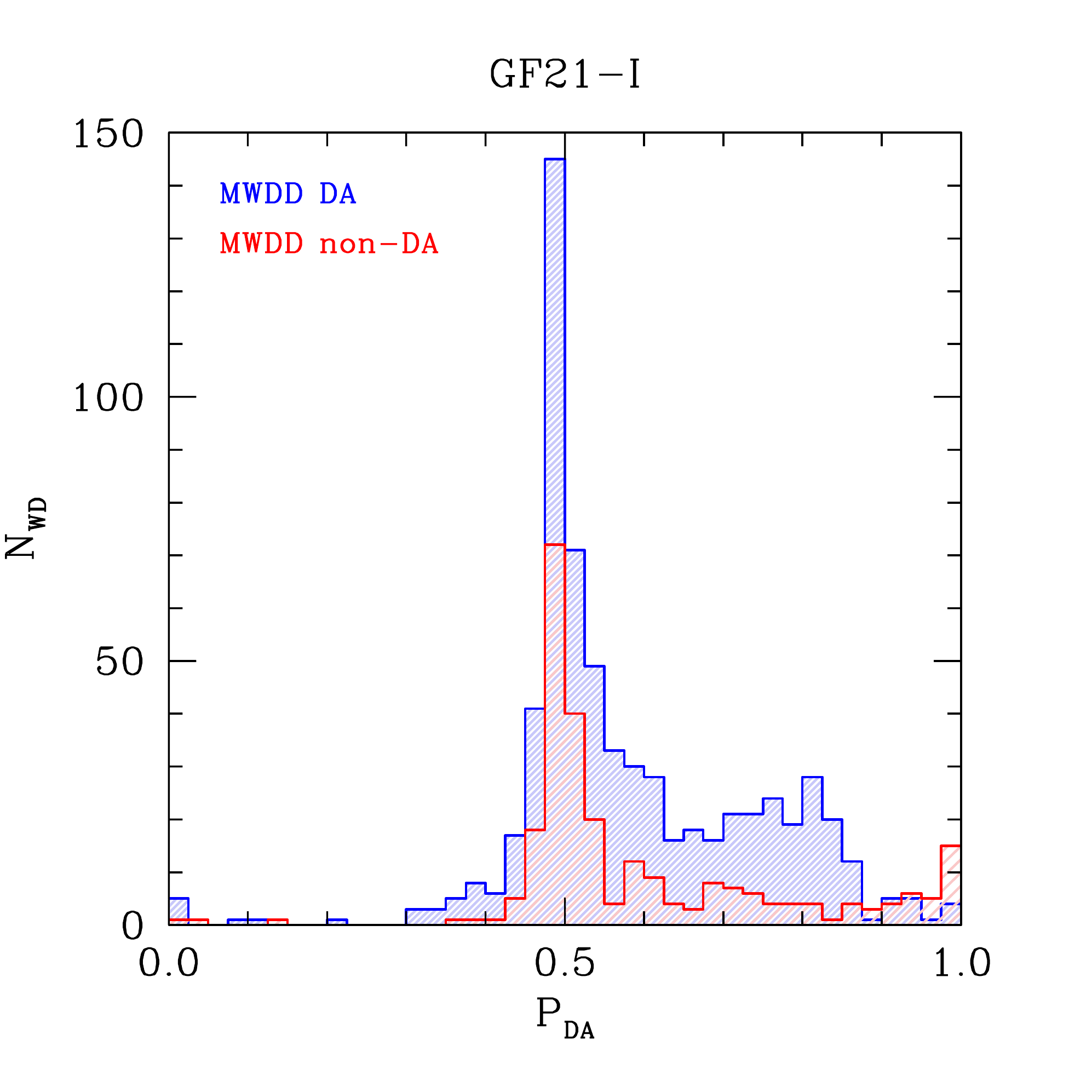}
  \includegraphics[width=0.32\textwidth,trim=2 10 20 20, clip]{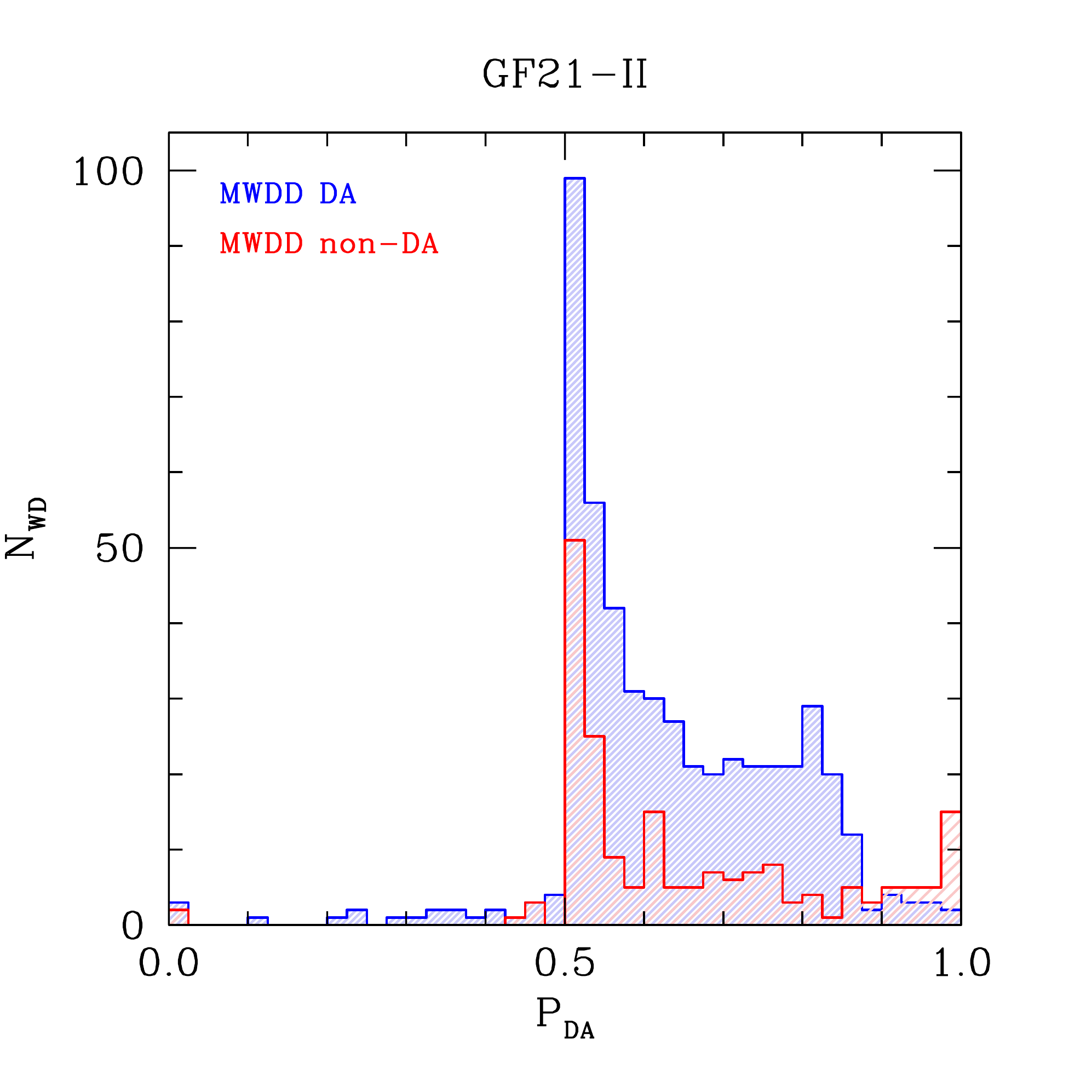}
  \includegraphics[width=0.32\textwidth,trim=2 10 20 20, clip]{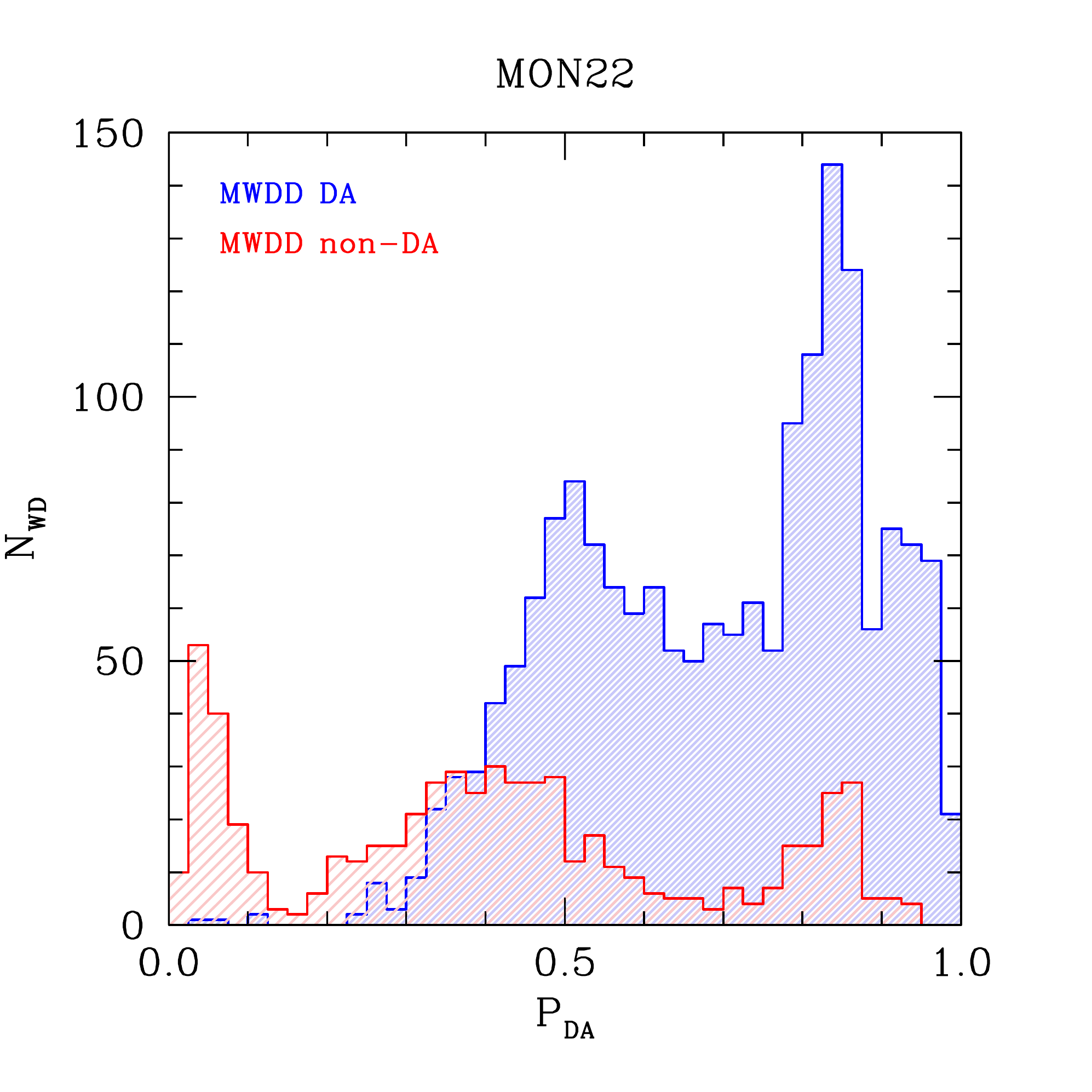}
\caption{Probability distribution of being DA for the different estimators analyzed in this work for the MWDD validating sample.}
\label{f:pdis_models}
\end{figure*}

\subsubsection{The GF21-I and GF21-II estimators}

The analysis of the {\it Gaia} Early Data Release 3 white dwarf population performed by \cite{GentileFusillo21} also provides a fitting of different theoretical atmospheric models to the three {\it Gaia} passbands: $G$, $G_{\rm RP}$, and $G_{\rm BP}$. Three models were used: one for hydrogen-pure atmospheres of \cite{Tremblay2011} with Lyman $\alpha$ opacity of \cite{Kowalski10}, another for pure-helium atmospheres based on \cite{Bergeron2011} models, and the last one for mixed H/He atmospheres with a ratio $\log N(H)/\log N(He)=10^{-5}$ based on models of \cite{Tremblay2014} and \cite{McCleery2020}. All the three models cover the range of effective temperatures and surface gravities of the white dwarfs analyzed here \citep[see][and reference therein for further details]{GentileFusillo21}.

For each atmospheric model -- hydrogen-pure, helium-pure, and mixed atmospheres -- a $\chi^2$-value, i.e., $\chi^2_{\rm H-pure}$, $\chi^2_{\rm He-pure}$ and $\chi^2_{\rm mixed}$, respectively, was provided in \cite{GentileFusillo21}. From the first two $\chi^2$-values, $\chi^2_{\rm H-pure}$ and $\chi^2_{\rm He-pure}$ ($\chi^2_{\rm DA}$ and $\chi^2_{\rm non-DA}$, respectively), we build from eq.\,\ref{e:probDA} our fourth DA probability estimator called GF21-I. Similarly, from the first and third $\chi^2$-values, $\chi^2_{\rm H-pure}$ and $\chi^2_{\rm mixed}$ ($\chi^2_{\rm DA}$ and $\chi^2_{\rm non-DA}$, respectively), we generated our fifth estimator named GF21-II.

\subsubsection{The MON22 estimator}

All except one of the estimators analyzed in this work follow eq.\,\ref{e:probDA}. The exception is the sixth estimator used in this work. It is represented by the probability of being a DA estimated by \cite{Montegriffo2022}. In that case, that probability was directly derived from the application of a Random Forest algorithm to the J-PAS synthetic photometry derived from {\it Gaia}-DR3 spectra.

\subsection{Validating sample: the Montreal White Dwarf Database}

MWDD\footnote{\url{https://www.montrealwhitedwarfdatabase.org/}} is an open access tool containing spectroscopically classified white dwarfs published in the literature. At the time of writing this article the database contained 68\,364 objects, 3098 of which are within 100\,pc from the Sun. From them, we rejected those which are binaries, those with circumstellar disc, and those with tentative spectral clasification. This reduced the sample to 2886 sources. We used this data base to construct a validating sample. 

It is worth saying here that our white dwarf atmospheric models cover a different range of temperatures, i.e., down to 3000\,K for DA and down to 5500\,K for non-DA (see Set.\,\ref{s:models}). To ensure a proper classification of the {\it Gaia} sample we introduced a colour cut. Given that a typical $\sim$\,0.6\,\msun\ white dwarf with an effective temperature hotter than 5500\,K has $G_{\rm BP}-G_{\rm RP}$\,$<$\,0.86 mag, we restricted our subsequent analysis to objects in this range of {\it Gaia} colour.

Among the 2886 selected white dwarfs within 100\,pc with spectral clasification in the MWDD, there are 2400 with $G_{\rm BP}$\,$-$\,$G_{\rm RP}$\,$<$\,0.86\,mag, that we adopted as a validating sample, represents $\sim$\,30\% of our 100\,pc white dwarf sample in this colour range. In particular, the MWDD validating sample contains 1789 white dwarfs classified as DAs, and 611 white dwarfs of other spectral types DB, DC, DQ, DZ, and similar types, that we assigned to the non-DA group. Thus, the MWDD sample has a ratio of DA to non-DA of 74:26 with a significant fraction of the different spectral types. This fact, together with the aforementioned 30\% cross-match with our catalogue, places the MWDD as an excellent validating sample for our spectral estimator tests.

\subsection{Results: confusion matrices and scores}

\begin{figure*}
\centering
  \includegraphics[width=0.33\textwidth,trim=8 10 10 20, clip]{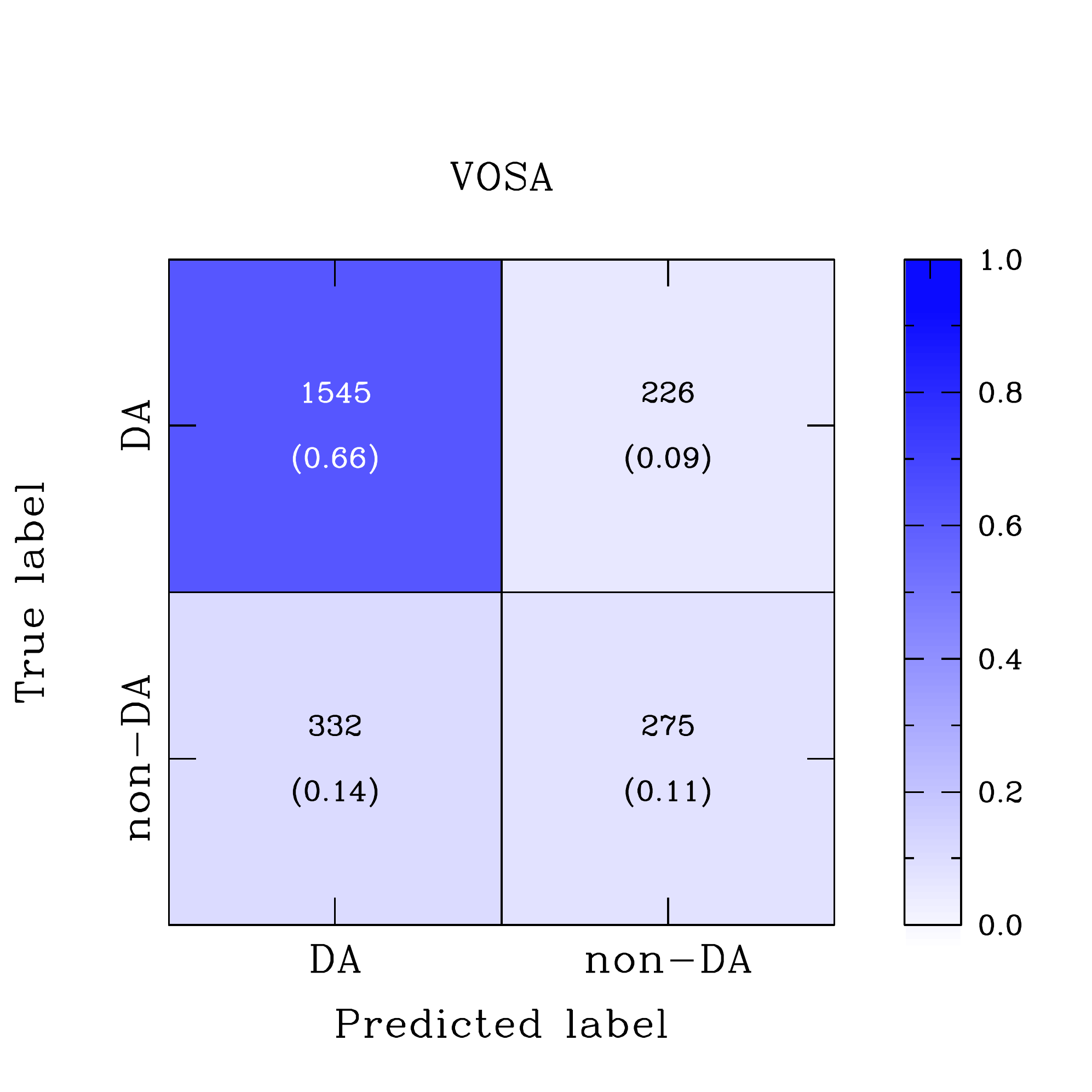}
  \includegraphics[width=0.33\textwidth,trim=8 10 10 20, clip]{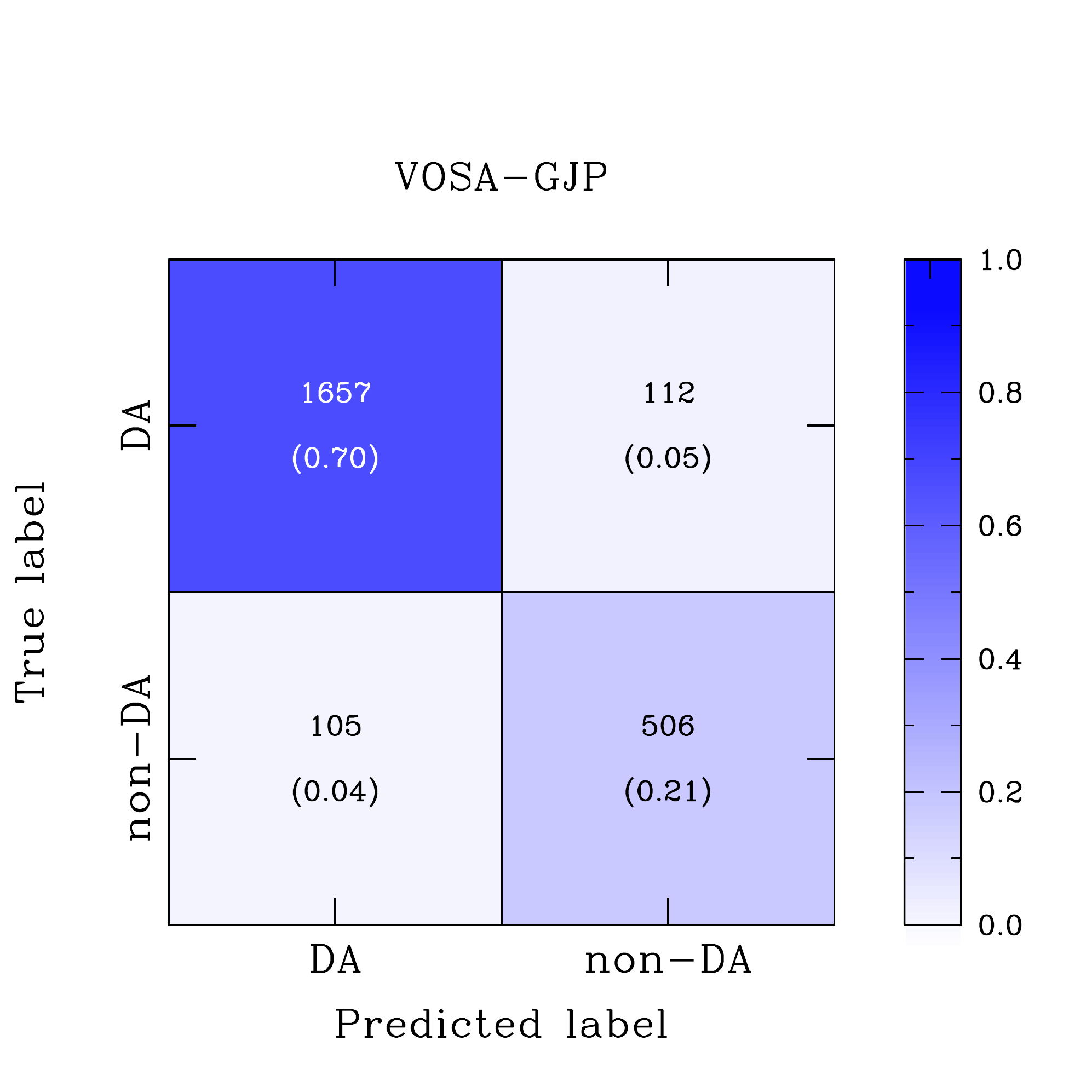}
  \includegraphics[width=0.33\textwidth,trim=8 10 10 20, clip]{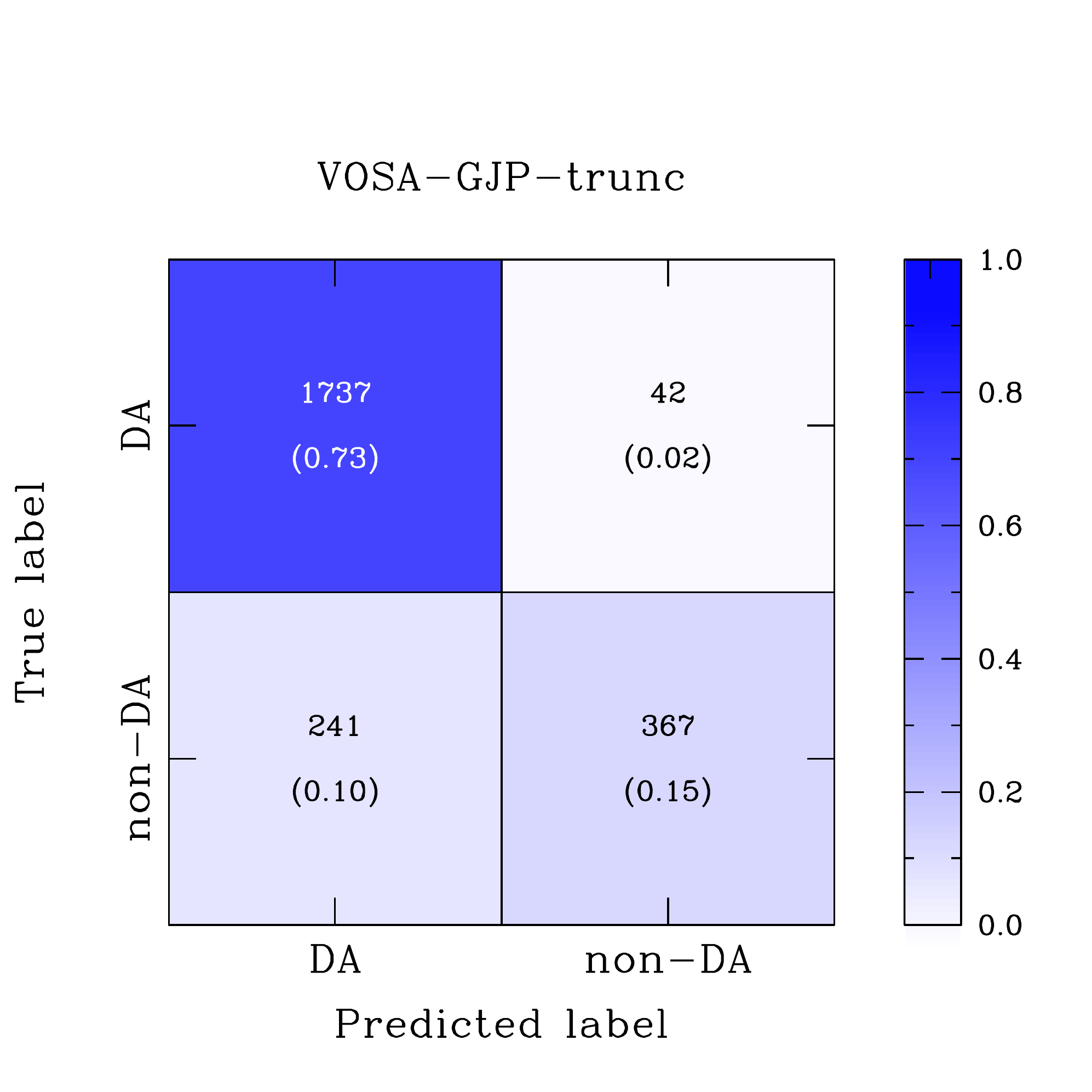}
       \vskip 0.3cm
  \includegraphics[width=0.33\textwidth,trim=8 10 10 50, clip]{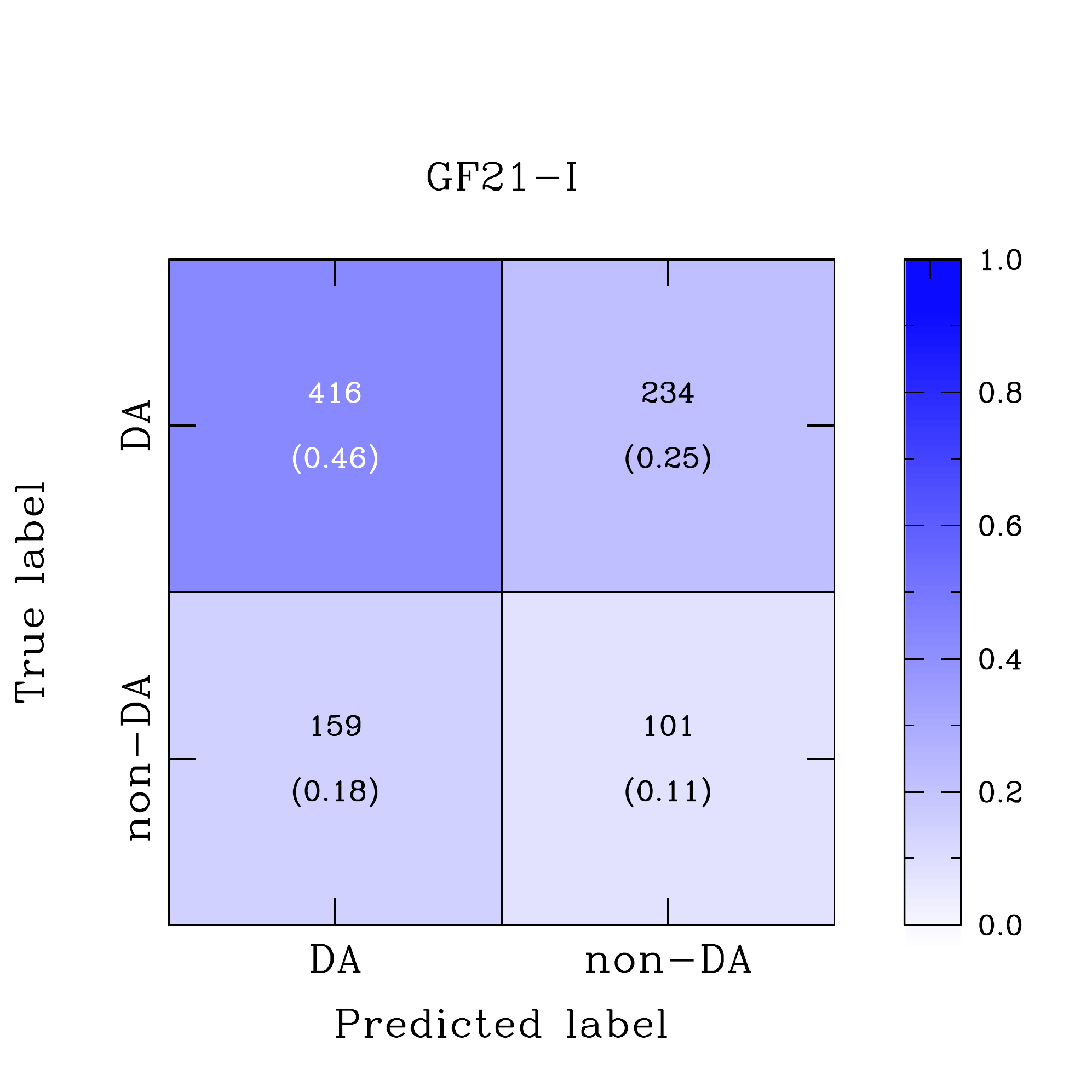}
  \includegraphics[width=0.33\textwidth,trim=8 10 10 50, clip]{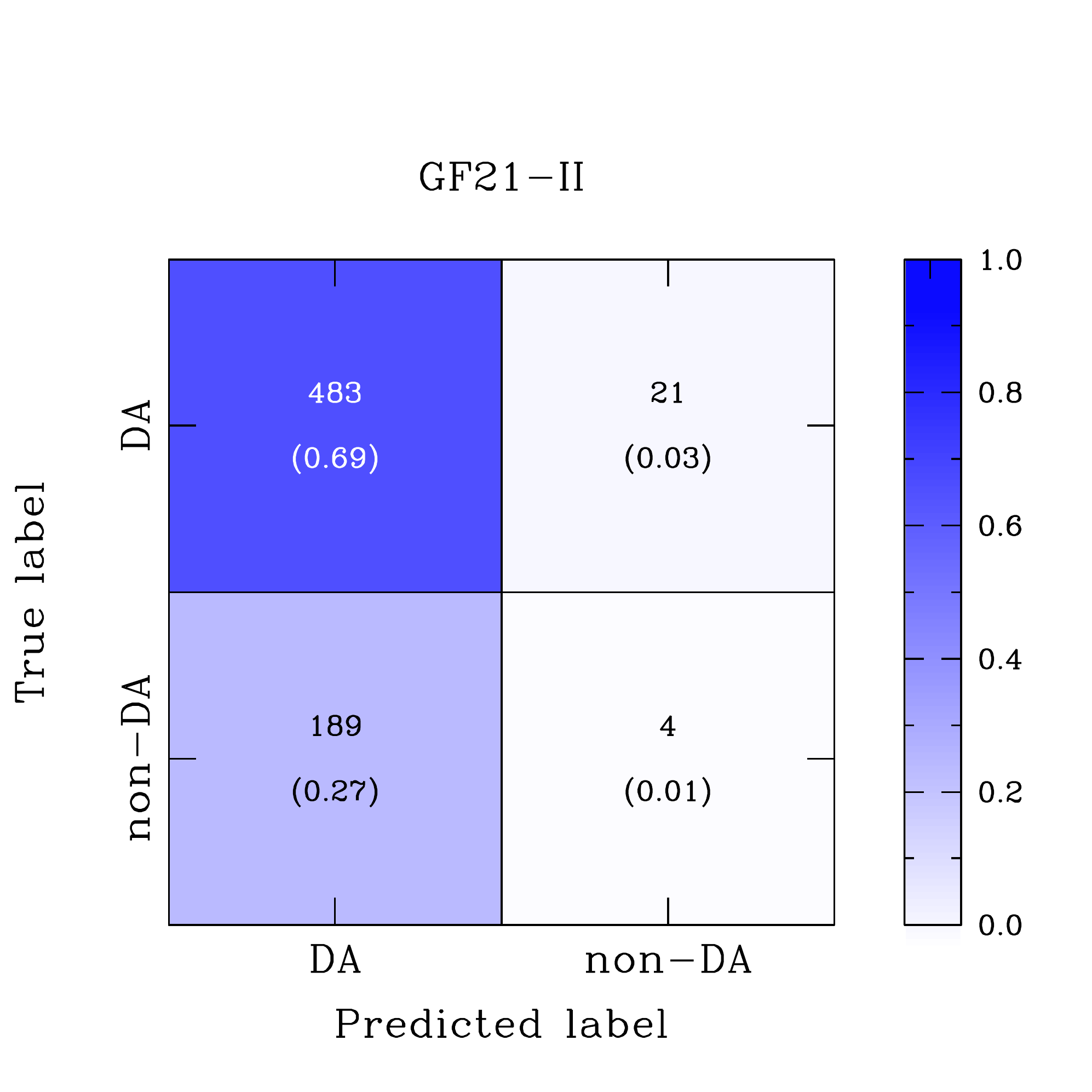}
  \includegraphics[width=0.33\textwidth,trim=8 10 10 50, clip]{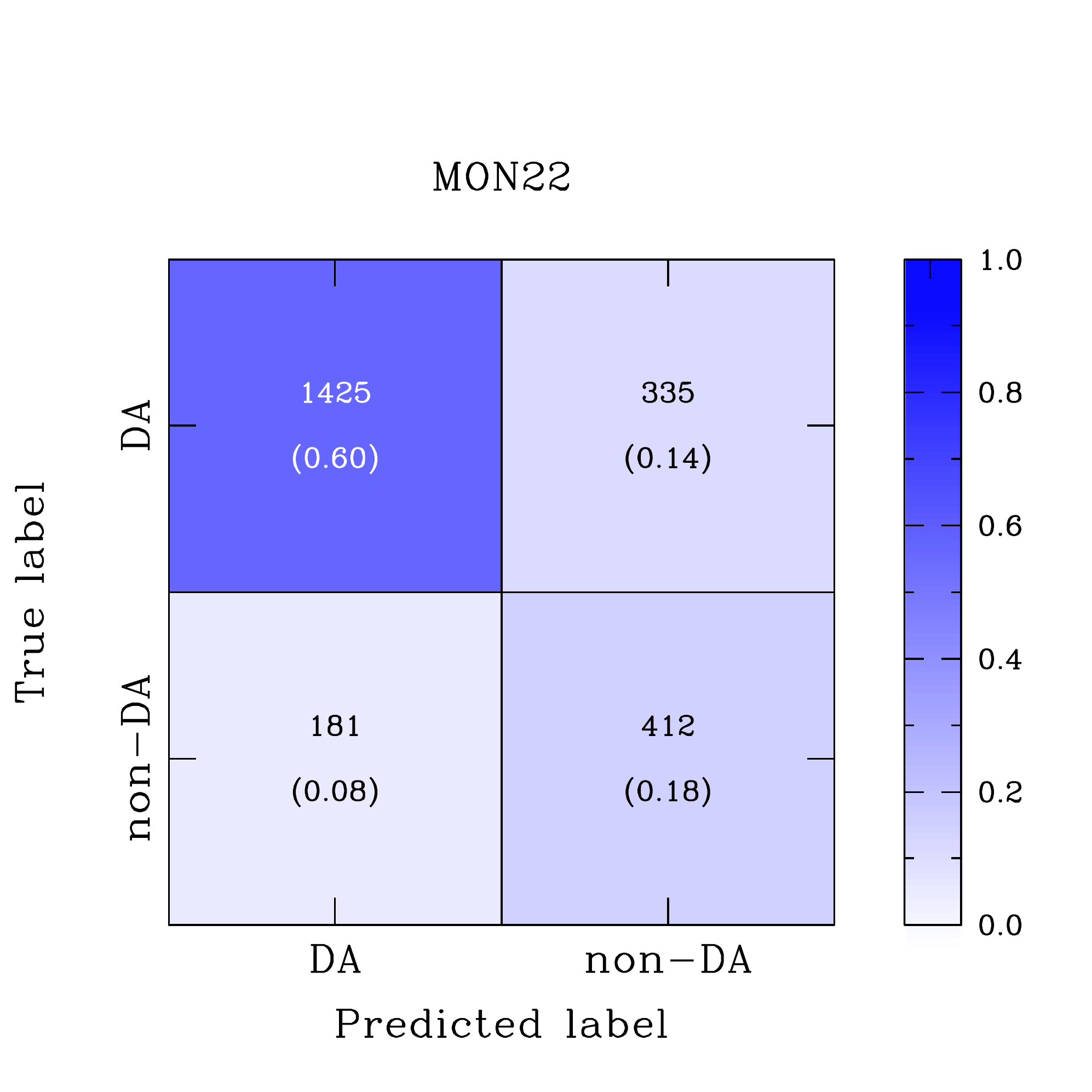}
  \caption{Confusion matrices for the different estimators analyzed in this work. Displayed values correspond to the total number of objects, while in brackets the normalized value with respect to the total population. Colour is scaled proportional to the total number of objects considered}
  \label{f:cm_models}
\end{figure*}

In Fig.\,\ref{f:pdis_models}, we show the probability distribution of being DA for the set of estimators under study. The ordinate axis represents the number of white dwarfs, while blue and red histograms correspond to white dwarfs belonging to the DA and non-DA classes, respectively, of the MWDD validating sample. A first glance to the distributions reveals that the pair of estimators VOSA-GJP and VOSA-GJP-trunc (top-middle and top-right panels, respectively) have a similar behaviour, indicating an excellent capability for disentangling DA from non-DA objects. In the case of the VOSA and MON22 estimators (top-left and bottom-right panels, respectively), the general behaviour is also acceptable. However, the fraction of non-DA white dwarfs misclassified as DAs is larger than in the two previous estimators. The last pair of estimators, GF21-I and GF21-II (bottom-left and bottom-middle panels, respectively), seem to have a similar behaviour, correctly identifying practically all DA white dwarfs, but failing in classifying most of the non-DA objects.

The previous analysis can be quantified by means of the confusion matrix, which represents the correlation between the true label object (in our case the MWDD DA or non-DA label; rows) as a function of the predicted label (the DA or non-DA estimator label; columns). The obtained results are shown in Fig.\,\ref{f:cm_models}, where the displayed values correspond to the total number of objects, the normalized values (in brackets) for each class, and the colour is scaled proportional to the total number of objects considered. The ideal case would correspond to a diagonal matrix with a fraction of DA:non-DA of 74:26, that is the fraction presented in the MWDD validating sample. Our estimator VOSA-GJP actually reproduces this ideal case, showing an excellent performance: only 6\% 
of true-DA white dwarfs are misclassified as non-DA, while less than 17\% 
of true non-DAs are not properly identified. 

In Table \ref{t:models} we list the accuracy and F1 global scores, as well as the sensitivity (also named recall) and the precision for each of the analyzed estimators. As previously stated, the estimator VOSA-GJP presents an excellent performance in all the analyzed scores. Particularly good are the accuracy and the F1-score, 0.91 and 0.94 respectively, as well as the precision of 0.94, reaching these scores the highest values among all the estimators. Regarding the sensitivity, the VOSA-GJP-trunc and GF21-II estimators, with 0.98 and 0.95, respectively, present a slightly better performance than VOSA-GJP, 0.94. However, the better ability of these estimators to retrieve objects of a certain class is worsened by a substantial lower performance regarding the precision with which an object is identified. All in all, the VOSA-GJP can be considered as an excellent estimator for the DA and non-DA white dwarf identification.

\begin{table}
\caption{Summary of the performance score, sorted from the highest to the lowest accuracy, for the different estimators under study.}
\begin{tabular}{l c c c c }
\hline \hline
 Estimator & Accuracy & F1-Score & Sensitivity & Precision \\
 \hline
VOSA-GJP    & 0.91 & 0.94 & 0.94 & 0.94 \\
VOSA-GJP-trunc & 0.88 & 0.92 & 0.98 & 0.88 \\
MON22      & 0.78 & 0.85 & 0.81 & 0.89 \\
VOSA      & 0.76 & 0.85 & 0.87 & 0.82 \\
GF21-II     & 0.70 & 0.82 & 0.96 & 0.72 \\
GF21-I     & 0.57 & 0.68 & 0.64 & 0.72 \\
\hline \hline
 \end{tabular}
\label{t:models}
\end{table}

Finally, before analyzing the physical properties of the identified samples in the next section, a few more checks have been made. As previously stated, for our initial analysis we adopted a threshold value of $\eta$\,=\,0.5. Now, we left $\eta$ as a free parameter. For the VOSA-GPJ estimator, the value that maximizes the performance is still $\eta$\,$\cong$\,0.5. For the rest of estimators, slight variations around this value are found. However, in none of the cases the performance of the rest of estimators improved that of the VOSA-GPJ.

\section{The physical properties of DA and non-DA 100\,pc white dwarf population}
\label{s:phys-prop}

In this section, we analyze the physical properties of the 100\,pc white dwarf population once classified according to DA or non-DA spectral type by our estimator VOSA-GJP.

We recall here that our analysis is restricted to those white dwarfs with good model fit to the GJP SED and in the {\it Gaia} colour range $G_{\rm BP}-G_{\rm RP}$\,$<$\,0.86\,mag, which approximately corresponds to a typical $\sim$\,0.6\,\msun\ white dwarf with an effective temperature just above 5500\,K. Thus, we did not expect $BP$-band photometry to be affected by any bias in this colour range, so we use this more commun {\it Gaia} colour in the analysis hereafter.

\begin{figure}
  \includegraphics[width=0.47\textwidth,trim=15 80 40 50, clip]{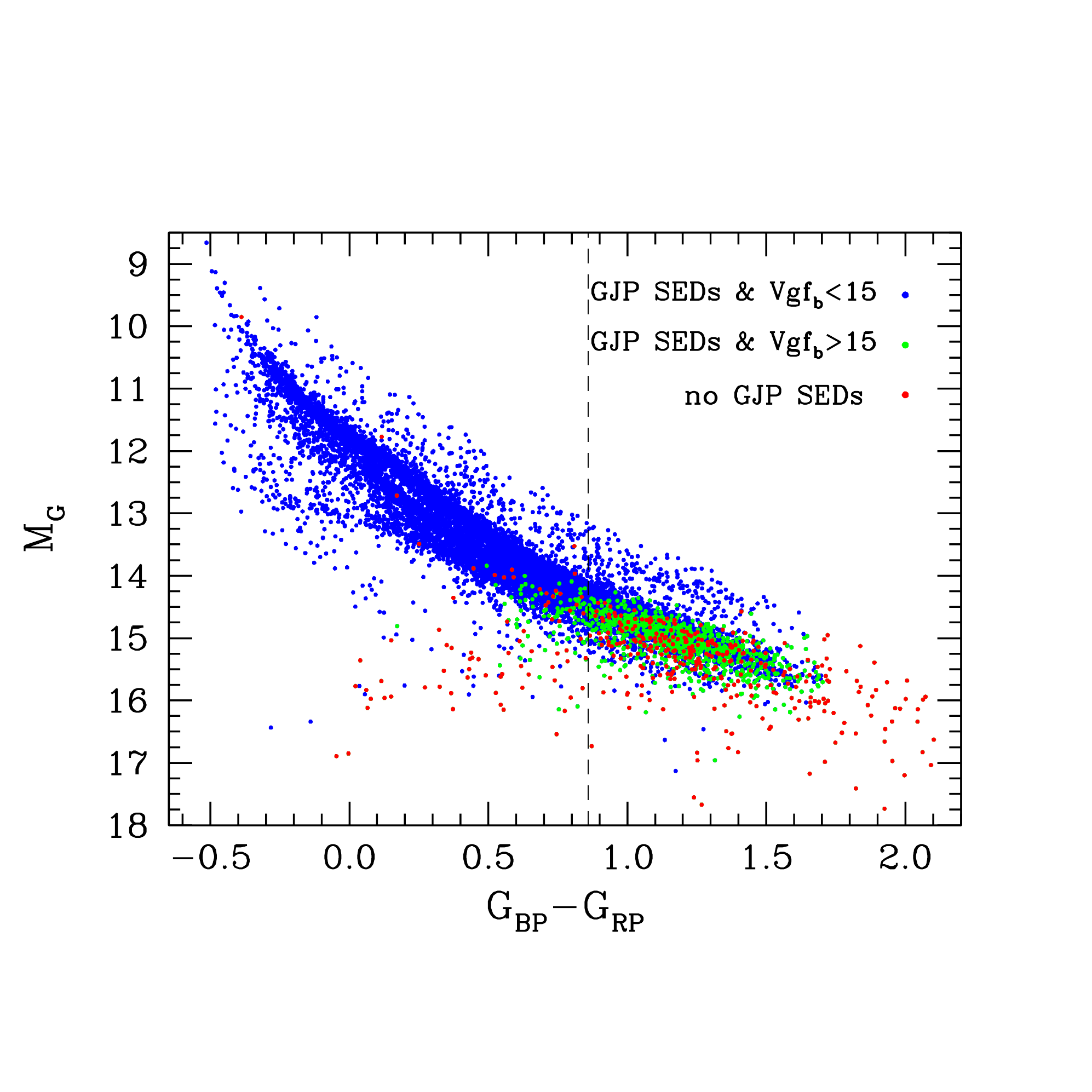} 
  \caption{{\it Gaia} Hertzsprung--Russell diagram of all {\it Gaia} sources considered in this work. Those with an available GJP SED and a reliable model fit (Vgf$_b$\,$<$\,15) are marked in blue. These objects in the range of colour $G_{\rm BP}-G_{\rm RP}$\,$<$\,0.86 (dashed line) are thoroughly analyzed in Section \ref{s:phys-prop}. Those sources with a bad fit are shonw in green, and those with no {\it Gaia} spectrum available in red. The total percentage of excluded objects represents only 1\% in the considered colour range.}
  \label{fig:HRD_SEDS}
\end{figure}

We show in Fig.\,\ref{fig:HRD_SEDS} the {\it Gaia} HR diagram for our 100-pc sample, where those objects with an available GJP SED obtained from the non-truncated {\it Gaia} spectra (see Sect.\,\ref{ss:SEDGaia}) and a reliable model fit (Vgf$_b$\,$<$\,15) are marked as blue dots. These objects which represent the 97\% of the entire 100\,pc white dwarf population. The remaining 7\% of objects have either a poor GJP SED fit (green dots) or even no {\it Gaia} spectrum is available (red dots). As it can be seen, the vast majority of them are faint and red objects. For the colour range considered in the analysis (left to the dashed line), the total number of objects classified is 8150, which represents the 99\% of the total population for that range of temperatures. This is up-to-date the most complete spectral type classified sample of white dwarfs.

\subsection{The Hertzsprung-Russell diagram for DA and non-DA white dwarfs}

\begin{figure*}
\centering
  \includegraphics[width=0.85\textwidth,trim=5 80 50 100, clip]{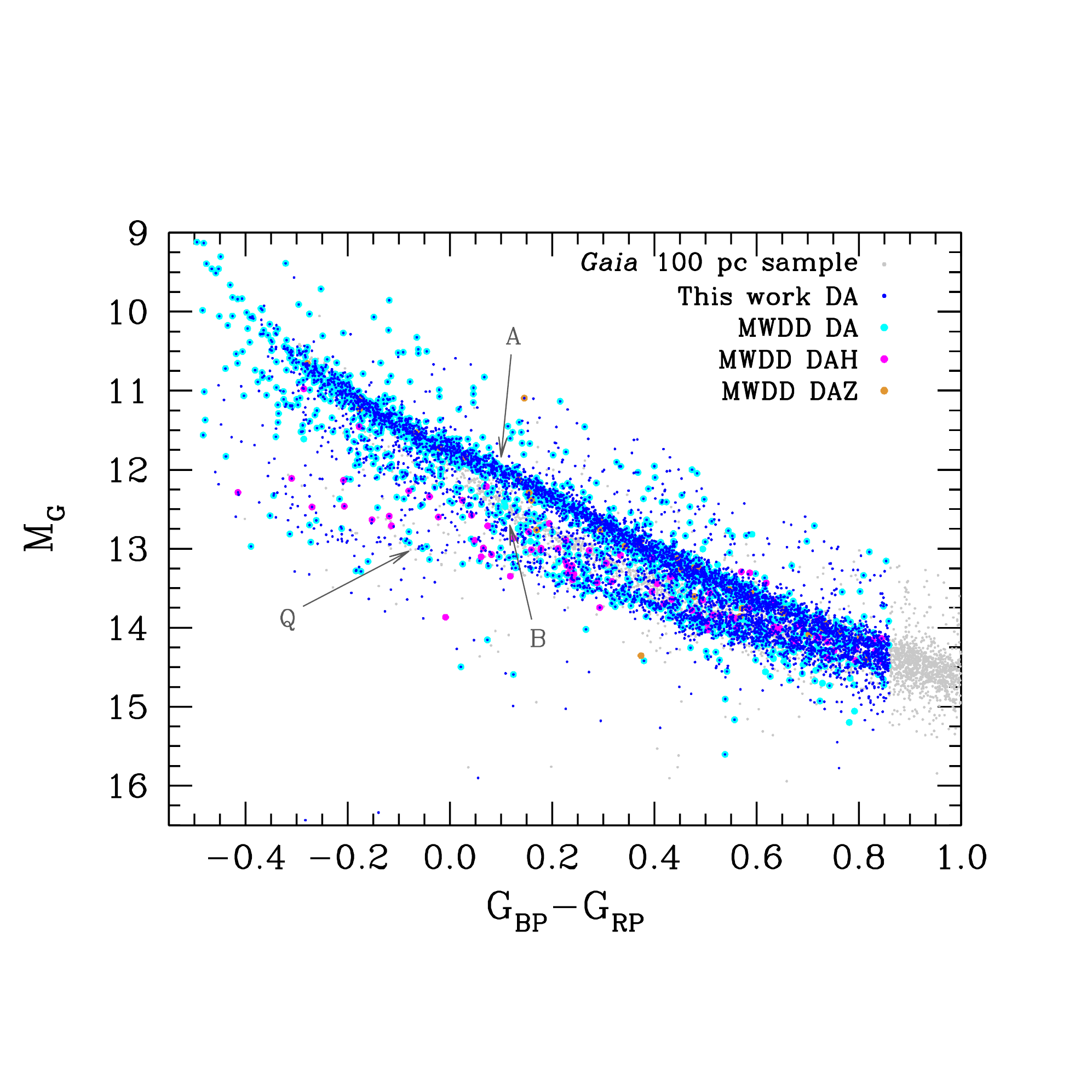}
  \includegraphics[width=0.85\textwidth,trim=5 80 50 100, clip]{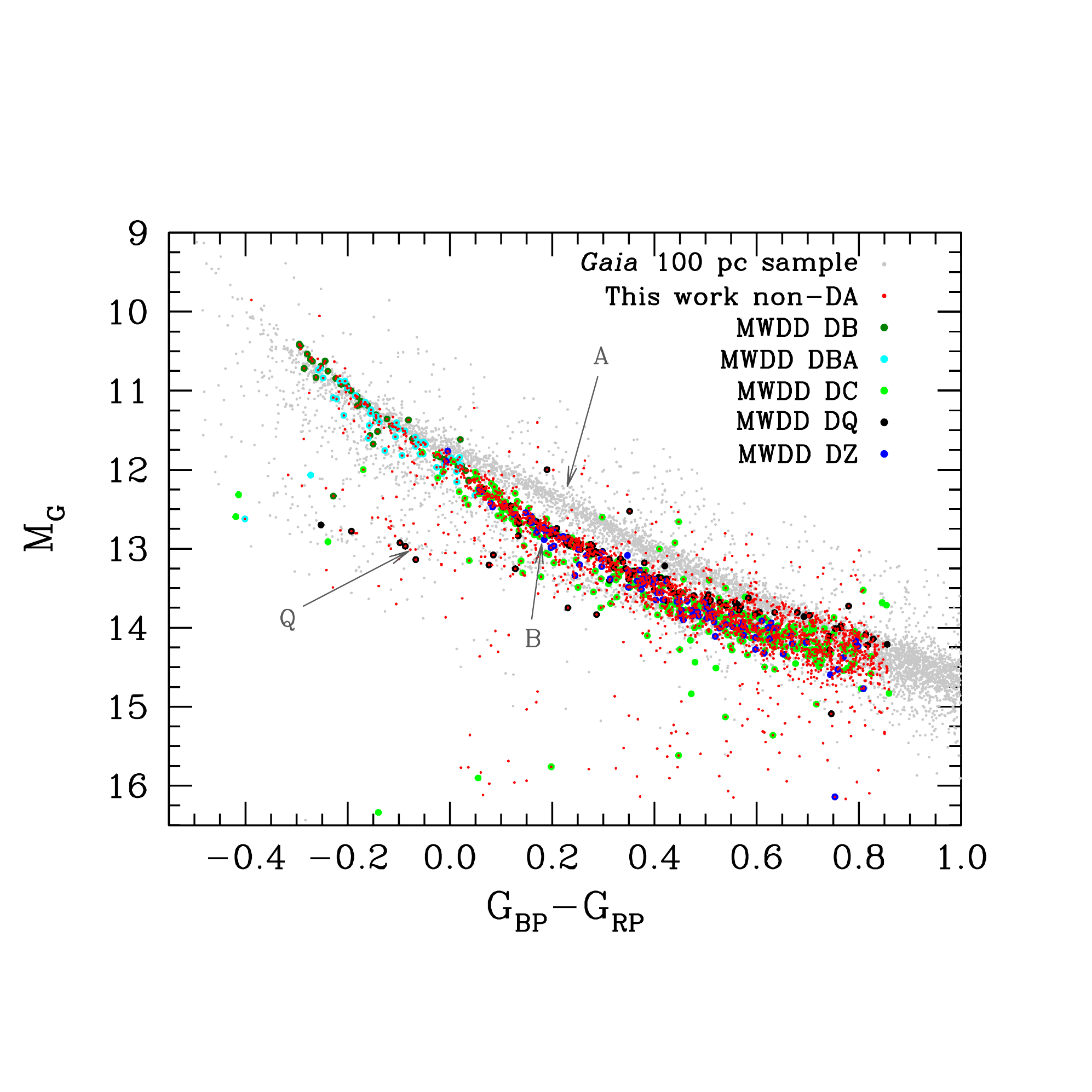}
  \caption{{\it Gaia} HR diagram for the DA (top panel; blue dots) and non-DA (bottom panel; red dots) white dwarfs classified in this work. Also plotted, for comparative purposes, our entire {\it Gaia} 100-pc sample (grey dots) and those classified in several spectral types in the MWDD. The A, B and Q branches are pointed for helping in the discussion. }
  \label{f:HRs}
\end{figure*}

In Fig.\,\ref{f:HRs}, we show the {\it Gaia} HR diagram for our sample of DA (top panel) and non-DA (bottom panel) white dwarfs. For comparative purposes, we have also plotted different spectral types of already classified white dwarfs according to the MWDD. The A, B and Q branches are also pointed \citep[see][]{Gaia2018}. 

A first glance at Fig.\,\ref{f:HRs} reveals an excellent agreement between the MWDD classified white dwarfs and those classified by our best estimator. This is not surprising since the accuracy of the VOSA-GJP estimator is larger than 90\%. Secondly, the loci in the colour-magnitude diagram of the DA and non-DA populations are clearly distinct. The characteristic bifurcation on the HR diagram has been put into manifest since the {\it Gaia}-DR2. However, the association of each of the two branches, A and B, to a pure-hydrogen or pure-helium atmosphere models, respectively, is a much debated problem \citep[e.g.][]{Jimenez-Esteban18,Bergeron2019}. In particular, if we assume that the B branch, within 0.1\,$<$\,$G_{\rm BP}-G_{\rm RP}$\,$<$\,0.5 mag, is formed by white dwarfs with pure-helium atmospheres, this would correspond to average masses larger than the canonical $0.6\,$\msun\ \citep[e.g.][]{Bergeron2019}. Several alternative explanations to avoid this problem have been proposed, such as mixed hydrogen-helium atmospheres, or spectral evolution from hydrogen to helium envelopes, among others \citep[e.g.][]{Bergeron2019, Ourique2020}. In any case, the characterization of the A and B branches has been limited to the spectroscopically identified white dwarfs in that region, leading to a large fraction of unidentified objects and, consequently, to ignore the real fraction of DA and non-DA in these branches. However, through our analysis, we can confirm that the B branch is mostly but not exclusively formed by non-DA white dwarfs, i.e. 65\% of non-DAs and 35\% of DAs, while the A branch is practically formed by DA white dwarfs (less than 6\% are non-DAs). Given that the proportion of sources in the A and B branches is 62 to 38\%, respectively, this implies that non-DA white dwarfs represent $\sim$\,25\% of the objects, considering both branches together.

A closer look to the HR diagrams reveals that DA white dwarfs are mostly following the main cooling track (A branch), corresponding to a canonical 0.6\,\msun\ white dwarf. However, as has been already mentioned, 18\% of the objects follows in parallel the main branch along the B branch. This bimodality was pointed out by \cite{ElBadry2018} and it was associated to a flattened in the initial-to-final mass relationship, which causes an overproduction of 0.8\,\msun\ white dwarfs. However, other possibilities may contribute to this issue, such as the contribution of white dwarf mergers \citep[e.g][]{Kilic2018}. Furthermore, a third branch, starting in the Q branch region, is growing in number of objects and joining the main branch at $G_{\rm BP}-G_{\rm RP}$\,$\sim$\,0.8\,mag. As we will analyze in Section \ref{ss:stellar}, the net effect is an increase of the average mass as objects are getting cooler. The combined effects of crystallization, $^{22}$Ne sedimentation, white dwarf mergers, and probably some other delaying physical cooling processes, may lead to the formation of this extended Q branch \citep[e.g.][]{Cheng2019,Tremblay2019,Kilic2020,Camisassa2021,2021ApJ...911L...5B}. 

Regarding the non-DA distribution, apart from the commented fact that they mostly are located in the B branch, we observe a widening of the track for the coolest objects, i.e., $G_{\rm BP}-G_{\rm RP}$\,$>$\,0.6\,mag. In this case, the reverse of the effect than in the DA sample is presented: a decrease in the average mass is expected for cooler non-DA objects. This effect, extended for cooler and less massive objects, was discussed in \cite{Bergeron2019}, although no final conclusion was reached. In any case, the true nature of non-DA objects that constitute the B branch, as well as the peculiarities in both diagrams already commented, are beyond the scope of the present work.

\subsection{Stellar parameter distributions}
\label{ss:stellar}

Relying on updated white dwarf evolutionary models from La Plata group and atmosphere models from \citet{Koester10}, we estimated stellar parameters for the white dwarfs with GJP SED, a reliable model fit (see Sect.\,\ref{ss:SEDGaia}), and $G_{\rm BP}-G_{\rm RP}$\,$<$\,0.86\,mag. We used different models depending on the spectral classification of the source (see Sect.\,\ref{s:class}). For DA white dwarfs, we used the cooling models of \citet{Althaus2013} for low-mass helium-core white dwarfs, the models of \citet{Camisassa2016} for average-mass carbon-oxygen-core white dwarfs, and the models of \citet{Camisassa2019} for ultra-massive oxygen-neon-core white dwarfs. For non-DA white dwarfs, we employed the hydrogen-deficient cooling models of \citet{Camisassa2017} for average-mass carbon-oxygen-core white dwarfs and \citet{Camisassa2019} for oxygen-neon ultra-massive hydrogen-deficient white dwarfs. All these models include realistic initial chemical profiles that are the result of the calculation of the progenitor evolution and consider all the relevant energy sources that govern their evolution. That is, they include neutrino loses, the gravothermal energy released by the ions, the energy released by slow $^{22}$Ne sedimentation and the energy released during the crystallization process, both as latent heat and due to the phase separation \citep[see][for details of its implementation]{Camisassa2022a}. We used the atmosphere models of \citet{Koester10} to turn these evolutionary models into the magnitudes in the {\it Gaia} passbands. For DA and non-DA white dwarfs we employed pure hydrogen and pure helium atmospheres, respectively. Considering the {\it Gaia} magnitude G and colour $G_{\rm BP}-G_{\rm RP}$ of each of the white dwarfs in our sample, we have interpolated in the theoretical models to obtain their masses and effective temperatures. We excluded all DA white dwarfs with estimated masses below 0.239\,M$_\odot$ and all non-DA white dwarfs with masses below 0.51\,M$_\odot$, to avoid extrapolation uncertainties.

\begin{figure*}
\centering
  \includegraphics[width=0.49\textwidth,trim=5 50 50 50, clip]{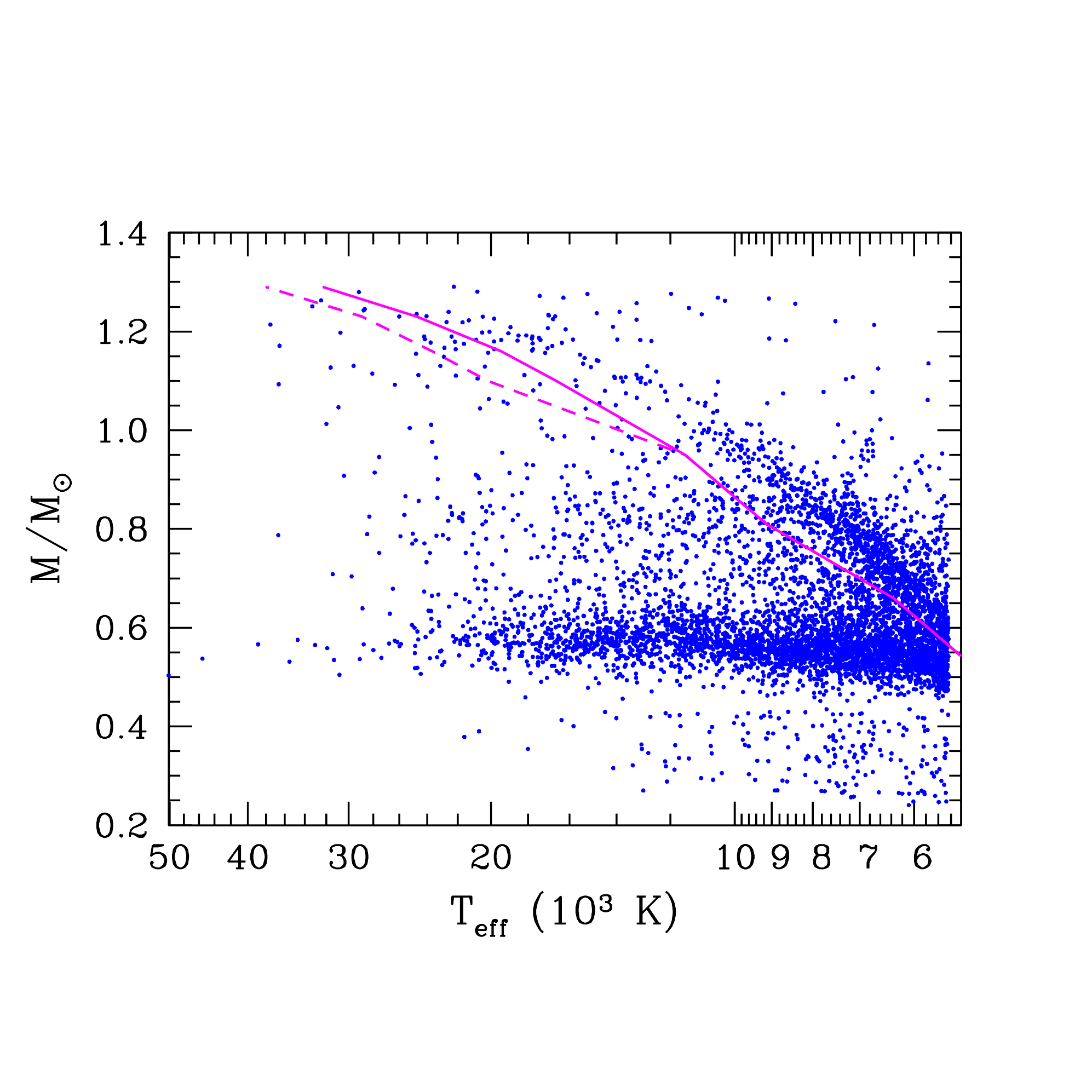}
  \includegraphics[width=0.49\textwidth,trim=5 50 50 50, clip]{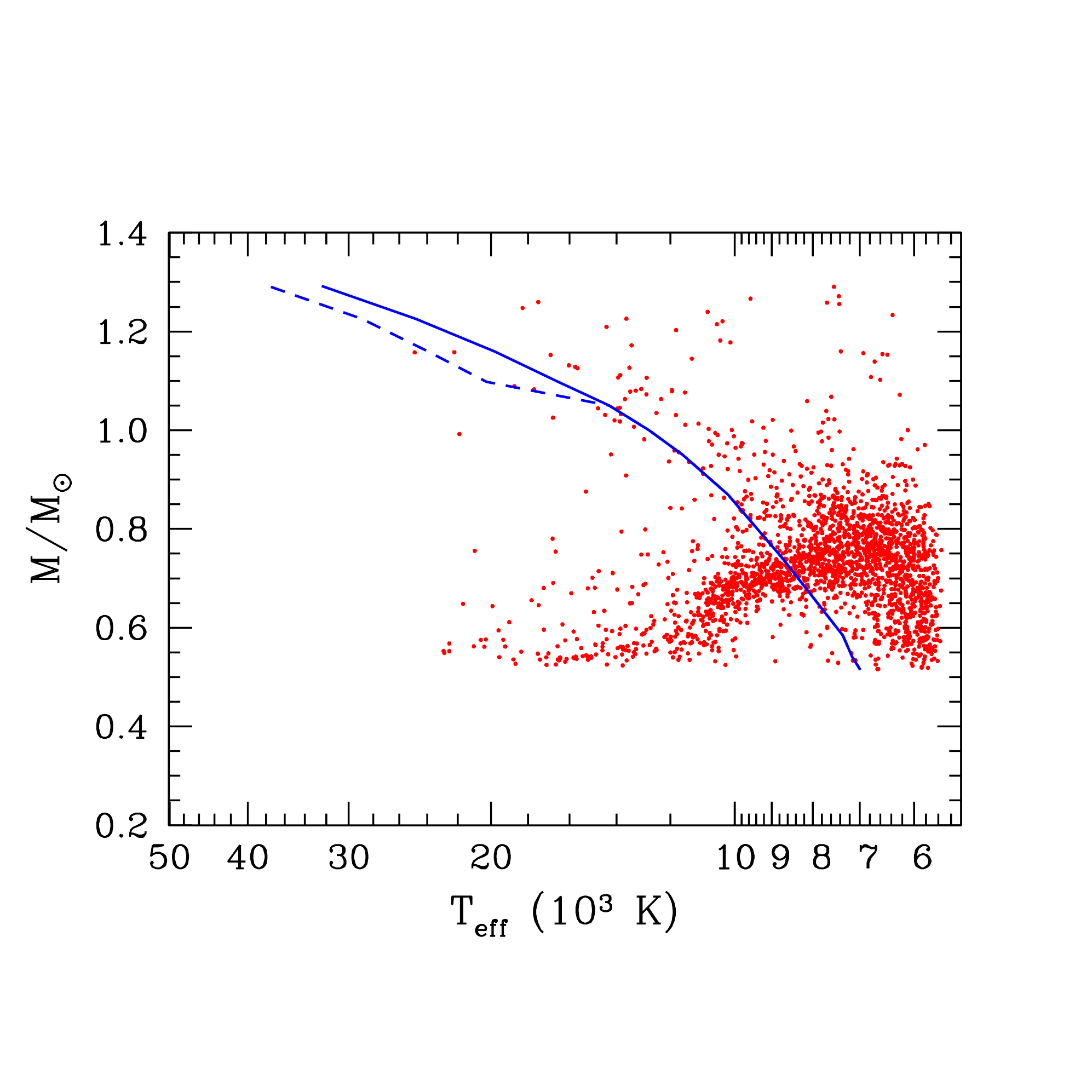}

  \caption{White dwarf masses vs effective temperature for our sample of DA (left panel) and non-DA (right panel). In addition, the crystallization onset assuming CO-core for all white dwarf (solid lines) or ONe-cores for masses above 1.05\,\msun\ (dashed lines) are plotted in magenta colour for DAs and in blue for non-DAs.}
  \label{f:massteff}
\end{figure*}

In Figure \ref{f:massteff}, we show the distribution of white dwarf masses versus effective temperature for our sample of DAs (left panel) and non-DAs (right panel). Also marked are the lines of crystallization onset assuming CO-core for all white dwarfs (solid lines) or adopting ONe-cores for masses above 1.05\,\msun\ (dotted lines). The crystallization onset was specifically calculated for DA (magenta lines) and non-DA (blue lines) white dwarfs. The distributions obtained reflect most of the peculiarities previously discussed regarding the {\it Gaia} HR diagram and already analyzed in similar diagrams by \citet[][]{Bergeron2019} and \cite{Kilic2020}. The bulk of the DAs is formed by objects of $\sim$\,0.6\,\msun\ uniformly distributed for the full range of temperatures. A secondary similar group, with masses around $\sim$\,0.8\,\msun\ but more diffuse can be intuited. Moreover, following the DA crystallization onset line, a group of massive and hot white dwarfs (starting with masses from $\sim$\,1.2-1.3\,\msun\ and $\sim$\,30\,000\,K), whose average mass is decreasing with the temperature is clearly visible. This last trend of stars corresponds to the Q branch and its extension, and they are related to the slow down on the white dwarf cooling rate due to the energy released by crystallization and by the sedimentation of $^{22}$Ne \citep{Tremblay2019,Camisassa2021}. The combined effect of these two later groups of stars for temperatures below 9000\,K is an increase on the canonical 0.6\,\msun\ average mass. Lastly, an increasing number of low-mass stars (due to double-degenerate or unresolved binaries) for lower temperatures is also observable. 

Regarding the non-DA distribution of masses and temperatures, we found a lack of $\sim$\,0.6\,\msun\ objects for temperatures between 7000\,$<$\,$T_{\rm eff}$\,$<$\,10\,000\,K. This effect, as pointed out in \cite{Bergeron2019}, can be avoided with the use of other atmospheric models such as mixed hydrogen-helium envelopes, instead of the pure-helium models we used. Likewise, the bulk of white dwarfs that have an average mass of $\sim$\,0.6\,\msun\ at $\sim$\,20\,000\,K, presents a smaller value of $\sim$\,0.55\,\msun\ at lower temperatures of $\sim$\,6000\,K. In any case, it is out of the scope of the present study to ascertain the ultimate atmospheric composition of these objects. Independently of this issue, the non-DA distribution also shows a clear increase in the number of massive objects for cooler temperatures. 

\begin{figure*}
\centering
\includegraphics[width=0.47\textwidth,angle=0,trim=1 10 50 10, clip]{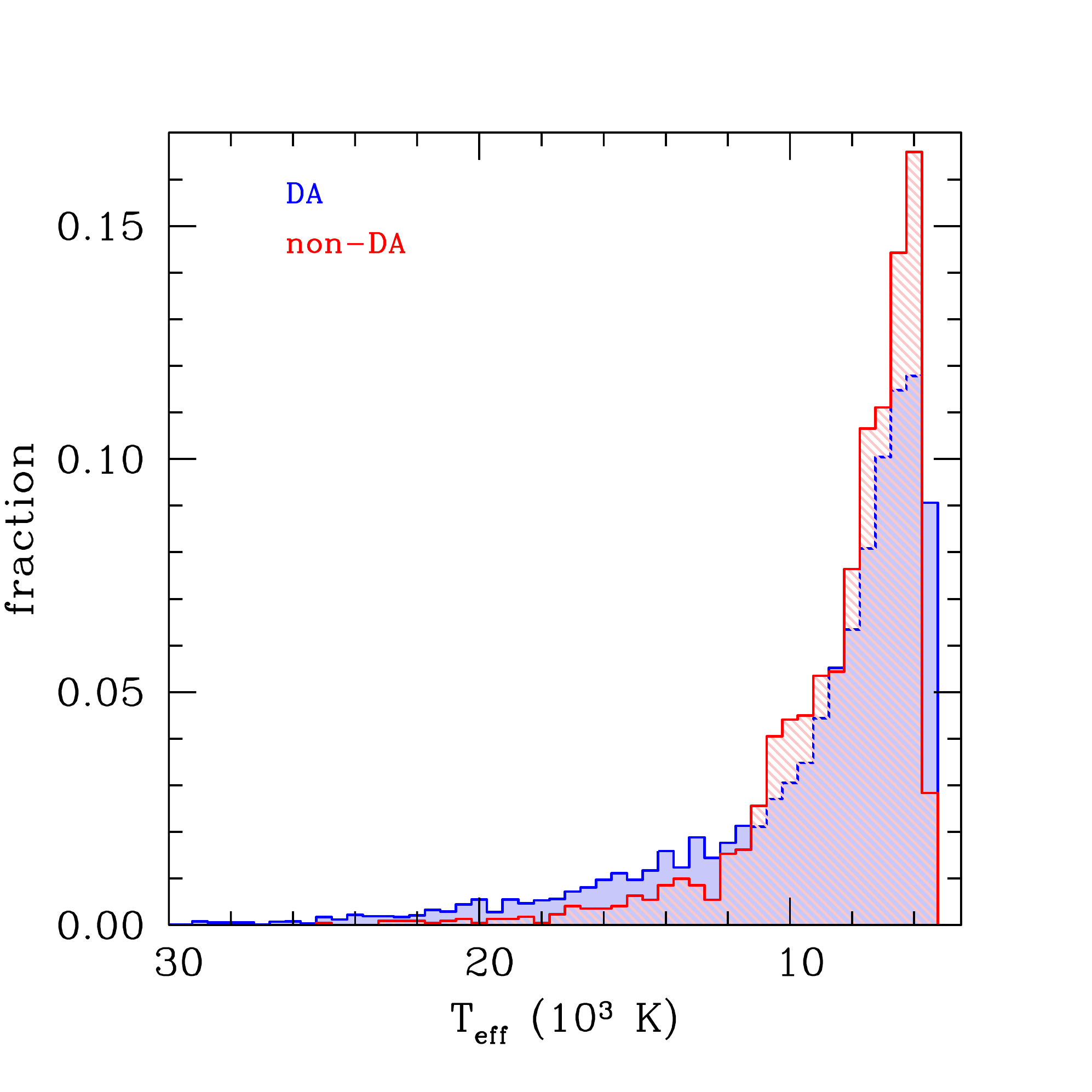}
\includegraphics[width=0.47\textwidth,angle=0,trim=1 10 50 10, clip]{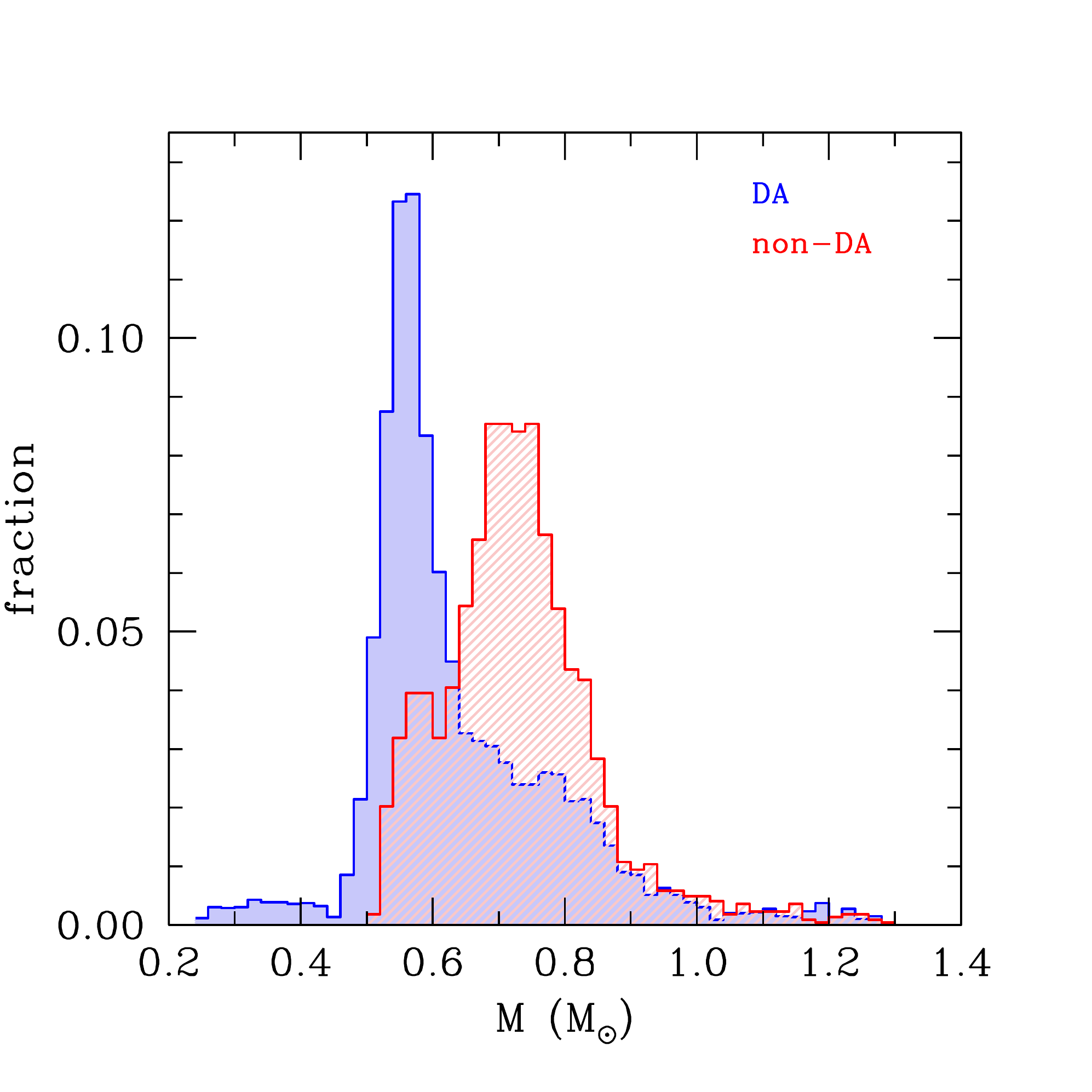}
  \caption{Distributions of effective temperature (left panel) and mass (right panel) for our sample of identified DA (blue histograms) and non-DA (red histogram). See text for details.}
  \label{f:parameters}
\end{figure*}

An individual analysis of the parameters is carried on in Fig.\,\ref{f:parameters}, where we show the frequency distribution of effective temperature (left panel) and mass (right panel) for our samples of identified DA (blue histogram) and non-DA (red histogram) white dwarfs. Roughly speaking, the distributions of effective temperatures seem to be similar for DA and non-DA white dwarfs. Although a deficit of non-DA objects is observable for hotter stars, $T_{\rm eff}$\,$\gappr$\,12\,000\,K, an excess occurs for cooler temperatures, $T_{\rm eff}$\,$\lappr$\,9000\,K. We further investigate this issue in the next section. 

Regarding the mass distribution, the discrepancies are notorious. The DA distribution shows the canonical peak at $\cong$\,0.58\,\msun, a bump extended up to 0.8\,\msun, and a small fraction of low-mass (non-single origin) white dwarfs. These trends are in agreement with previous reported works \citep[e.g.][]{Jimenez-Esteban18,Kilic2020}. With regard to the non-DA distribution, a clear bimodality with peaks at $\sim$\,0.6\,\msun\ and $\sim$\,0.76\,\msun\ is present. As stated, this second peak is a consequence of the bifurcation in the HR diagram and the use of pure-helium atmospheres to derive its mass. Otherwise, for instance, in the case of mixed hydrogen-helium atmospheres, this second peak can eventually be removed, or at least reduced. However, if we adopt a flattened initial-to-final mass relationship as in the case of the DA population, a similar peak around 0.8\,\msun\ should be expected in the non-DA mass distribution. Under this hypothesis, and assuming the same ratio between the peaks at 0.6 and 0.8\,\msun\ as in the DA case (i.e., 80:20 respectively, see Fig.\,\ref{f:parameters} right panel), we would expect that 20\% of the non-DA white dwarfs in the B branch to have masses in the range of $\sim$\,0.8\,\msun.

\begin{figure*}
\centering 
\includegraphics[width=0.45\textwidth,trim=60 10 100 30, clip]{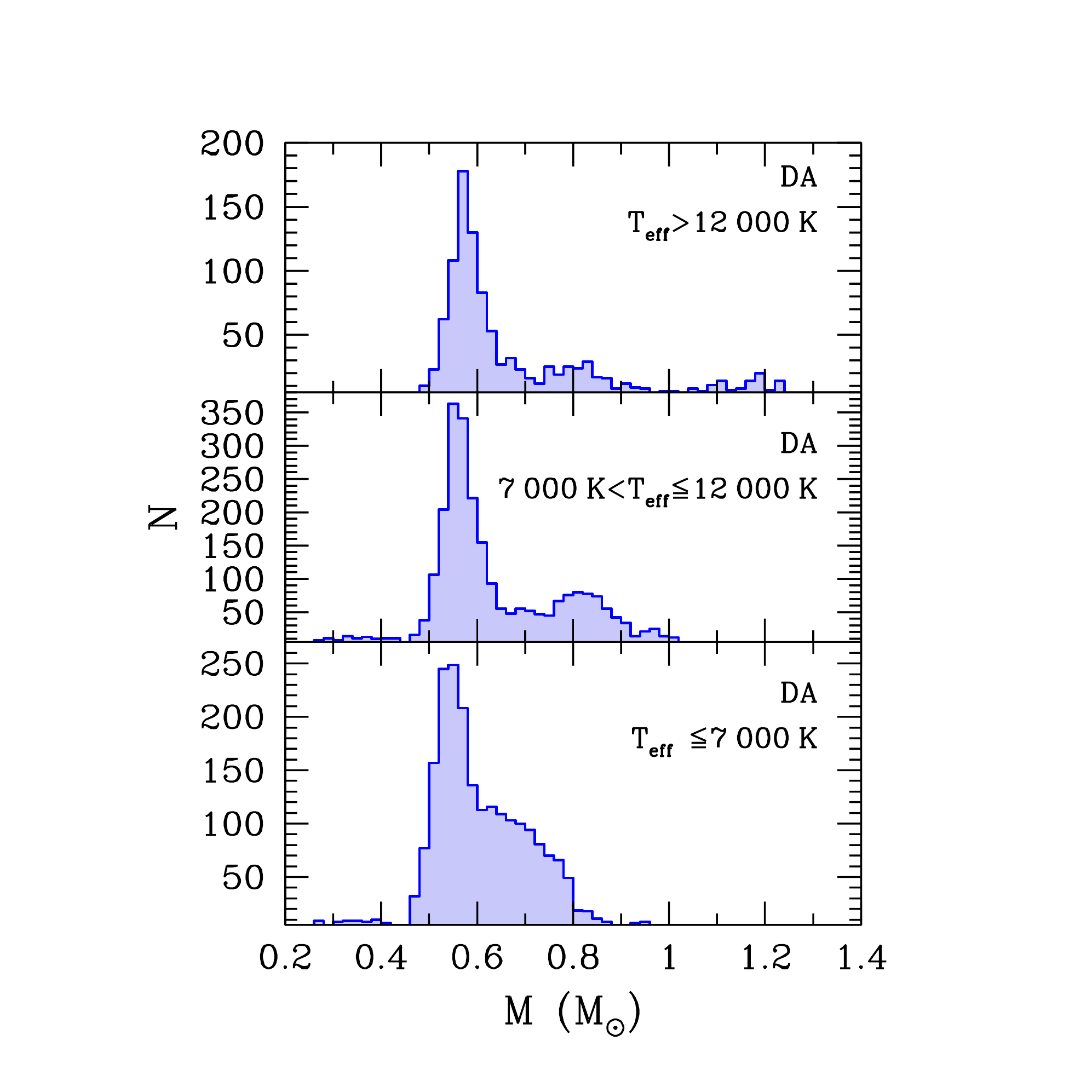}
\includegraphics[width=0.45\textwidth,trim=60 10 100 30, clip]{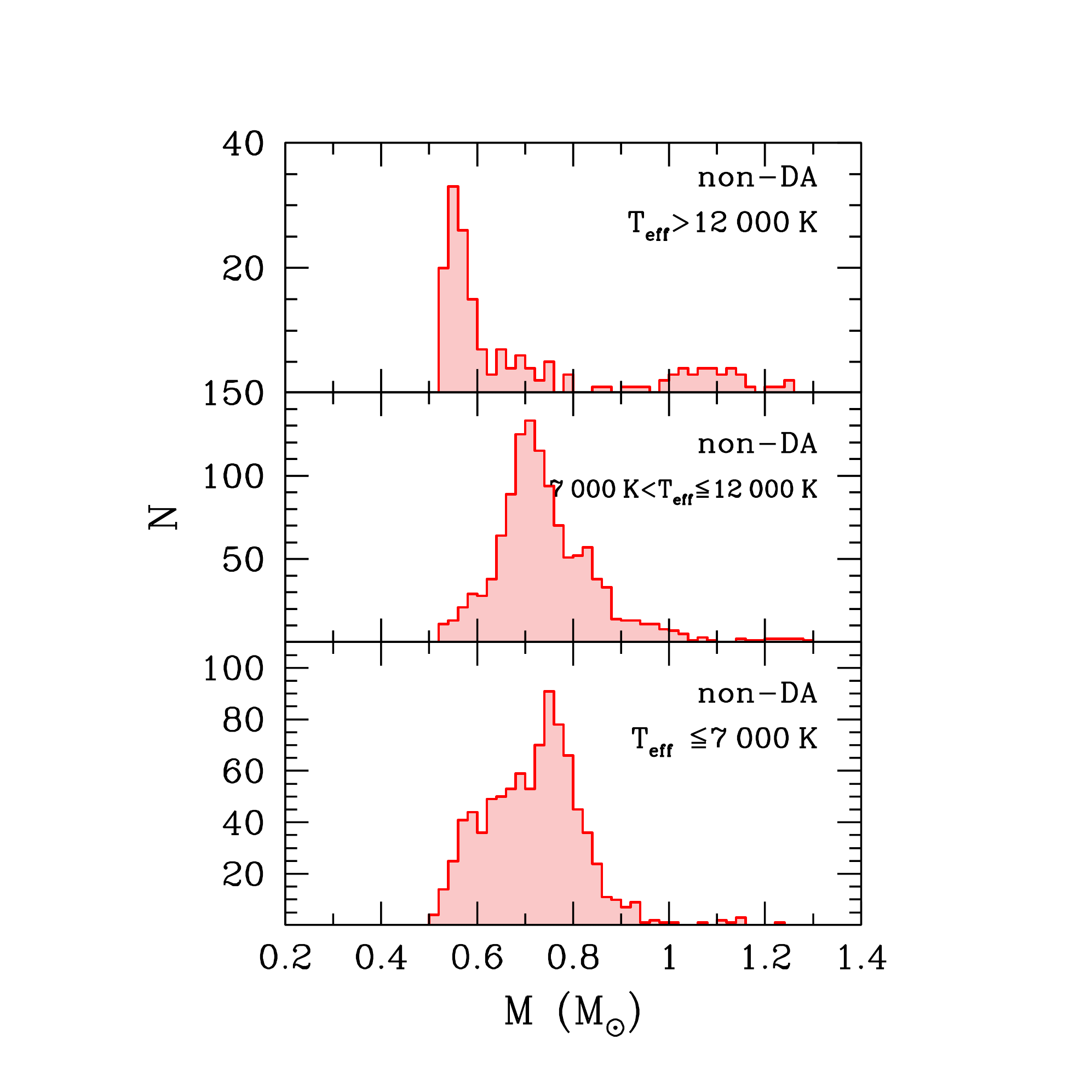}
  \caption{Distribution of mass for our sample of identified DA (left panels) and non-DA (right panels) in different range of temperatures.}
  \label{f:massdisteff}
\end{figure*}

Further information can be retrieved when the mass distribution is depicted as a function of the effective temperature. This is done in Fig.\,\ref{f:massdisteff}, where we plot the DA (left panels) and non-DA (right panels) mass distribution for hot ($T_{\rm eff}$\,$>$12\,000\,K; top panels), medium (7000\,$<$\,$T_{\rm eff}$\,$\leq$\,12\,000\,K; middle panels), and cool ($T_{\rm eff}$\,$\leq$\,7000\,K; bottom panels) white dwarfs. These temperature ranges approximately correspond, respectively, to the regions before, during, and after the bifurcation in the \G HR diagram. 

The DA mass distribution for hotter white dwarfs presents the canonical peak at $\cong$\,0.58\,\msun\ and an extended tail for massive white dwarfs. However, for $T_{\rm eff}$\,$<$\,12\,000\,K practically no objects more massive than $1.0\,$\msun\ are found. Two factors should be taken into account. On one hand, as it is seen in Fig.\,\ref{fig:HRD_SEDS}, the 1\% of sources excluded (red and green dots) in the analysed colour range have larger absolute magnitudes and redder colour. These objects typically correspond to cool massive white dwarfs. On the other hand, as pointed out by \cite{Kilic2020}, this absence of more massive stars can be also associated to the delay in the cooling process due to the effect of core crystallization, among others. Moreover, while the main peak remains almost constant regardless of the temperature range, the second one at $\sim$\,0.8\,\msun\ seems to be more prominent in the middle range of temperatures. Even more, for the coolest white dwarfs an extended tail from 0.6 to 0.8\,\msun\ appears. This feature is indicative that, even assuming the bimodality is caused by the flattened of the initial-final mass relationship, an additional process such as the contribution of the merger white dwarf population needs to be added to this scenario. Some of these features on the mass-temperature distribution have been partially guessed in previous works \citep[e.g.][]{Kepler2007,Tremblay2010,Rebassa2015a,Rebassa2015b,Tremblay2016}. However, it was not until the arrival of the data provided by the {\it Gaia}-DR3 that we had a full picture of the mass-temperature distribution of the white dwarf population.

In the case of the non-DA mass distribution, the dependence with the effective temperature is stronger than in the previous case. While the hottest non-DA white dwarfs peak at $\cong$\,0.58\,\msun\ with an extended tail up to massive white dwarfs, the non-DA mass distribution for medium temperatures peaks at $\cong$\,0.75\,\msun. We recall that this range of effective temperatures corresponds to the bifurcation and that the increase of the average mass is likely a consequence of the use of pure-helium atmospheres for white dwarfs in the B branch. In the case of the coolest non-DA white dwarfs, an extended fraction of low mass ($<$\,0.6\,\msun) white dwarfs appears. Nevertheless, all these features on the non-DA mass distribution should be taken with caution, given that we are assuming pure-helium atmospheres for all non-DA objects, and that a combination of different atmospheres such a mixed, carbon-contaminated, etc., is expected to be present in the observed sample.

\subsection{The ratio of DA to non-DA white dwarfs}

The ratio of DA to non-DA white dwarfs as a function of the effective temperature, also commonly referred to as the spectral evolution, is a capital tool for understanding the evolutionary processes in white dwarf atmospheres \citep[e.g.][and references therein]{Ourique2020,Cunningham2020}. Due to the diversity of selection effects, one of the major problems that previous studies had to face when building a detailed spectral evolution function was the incompleteness of the analyzed samples. However, based on the robustness of our estimator built from {\it Gaia} spectra, we could analyze with a highly degree of completeness our statistically significant 100\,pc white dwarf sample. That is, we can consider it as a volume limited sample, thus avoiding the need of any assumption on correcting factors. Furthermore, our analysis allowed to reconstruct the spectral evolution for a wide range of effective temperatures, from 25\,000 to 5500\,K.

\begin{figure*}
\centering
  \includegraphics[width=0.8\textwidth,trim=0 60 20 50, clip]{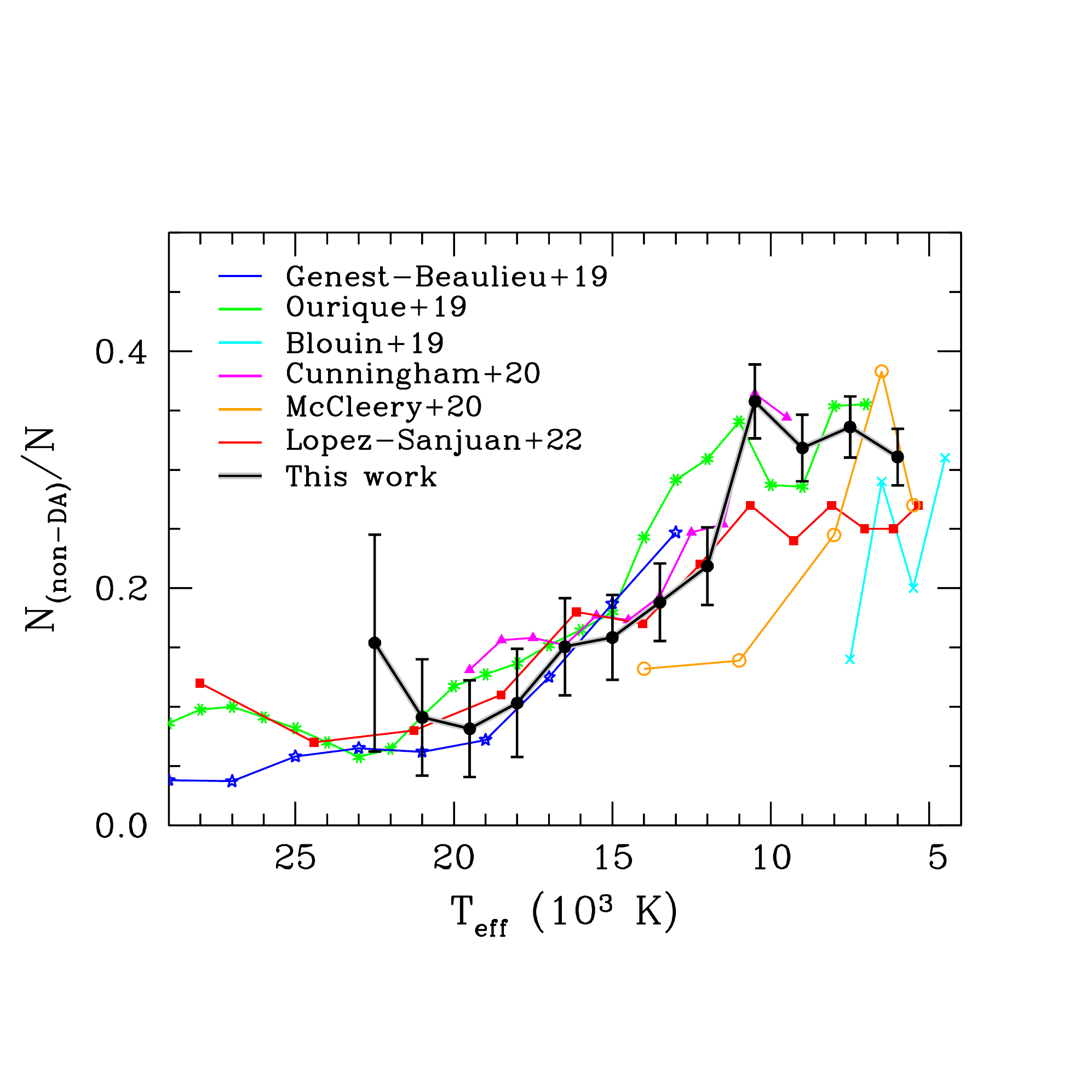}

  \caption{{\it Gaia} Ratio of DA vs non-DA for our sample of identified white dwarfs (black points and lines). Objects with $M_{\rm WD}$\,$<$\,0.51\,\msun\ has been disregarded to avoid contamination from unresolved binaries. For comparative purposes we also show spectral evolution distributions from other works.}
  \label{f:ratioDAnDA}
\end{figure*}

In Fig.\,\ref{f:ratioDAnDA}, we show the ratio of non-DA white dwarfs with respect to the total number of objects as a function of effective temperature (black line). We recall that our classification was done for the {\it Gaia} colour range $G_{\rm BP}-G_{\rm RP}$\,$<$\,0.86\,mag, corresponding to $T_{\rm eff}$\,$>$\,5500\,K and, in order to avoid the effect of unresolved binaries, we selected objects only with $M_{\rm WD}$\,$>$\,0.51\,\msun. Error bars were estimated as $\sigma_f=\sqrt{f\times (1-f)/N}$, where $f$ is the fraction of non-DA to the total number of objects, $N$, and weighting the contribution of each object according to the probability of being DA. For comparative purposes, we also show some other distributions recently published in the literature. 

The first characteristic of our spectral evolution function is the general smooth behaviour, a direct consequence of the high degree of completeness of our sample. Secondly, error bars are larger for hotter temperatures, given that our 100\,pc volume-limited sample is relatively small in comparison to magnitude-limited samples, and thus, the scarcity in hot objects. However, the higher resolution of our distribution for cooler temperatures is evident. Analyzing the particular details of our distribution, we discern some of the characteristics already presented in previous works. For instance, in the range of temperatures between $\sim$\,23\,000 and $\sim$\,13\,000\,K, our spectral distribution is nearly constant with an average ratio of $\sim$\,13\% and a slight growth up to 13\,000\,K. This behaviour is compatible with most of the distributions: \cite{Ourique2020}, \cite{Lopez-Sanjuan2022}, \cite{Cunningham2020}, and \cite{Genest2019}. At that point, there is an abrupt increase that reaches its maximum fraction, $f$\,$\cong$\,36\% at $T_{\rm eff}$\,$\sim$\,10\,500\,K, which corresponds to the bifurcation zone in the {\it Gaia} HR diagram. A similar high ratio is also found in \cite{Cunningham2020,Ourique2020}. In particular, our distribution perfectly resembles that of \cite{Cunningham2020}. For cooler temperatures, the fraction decreases down to $f$\,$\cong$\,0.31 for the coolest bin ($T_{\rm eff}$\,$\sim$\,5500\,K). A similar trend is also found by \cite{McCleery2020}. On the contrary, we have found no evidences of an increase ratio of non-DA to DA between 6250\,$\le$\,$T_{\rm eff}$\,$\le$\,7500\,K due to convective mixing, as predicted by \cite{Blouin2019}. However, as it is evident from the discrepancies among the distributions, any conclusion in the cooler range of temperatures should be taken with caution, as larger photometric errors and selection biases dominate in this region.


\section{Conclusions}
\label{conclusion}

An update of our 100\,pc white dwarf sample \citep{Jimenez-Esteban18} was performed, taking advantage of the new photometric and astrometric data together with the low resolution spectra provided by the {\it Gaia}-DR3. A total of 12\,718 white dwarfs configure our new catalogue, from which 12\,342 have {\it Gaia} spectra. This constitutes the largest nearly-complete volume-limited white dwarf sample available to date.

The use of automated algorithms for fitting and analysing SEDs, provided by the Spanish Virtual Observatory, and in particular VOSA, allowed us to extract the maximum information from the SED of the white dwarfs in our sample. In this sense, we built an estimator of the probability of being DA from the fitting of hydrogen-rich and nearly helium-pure atmosphere models to the synthetic J-PAS photometry derived from the {\it Gaia} spectra, when all the coefficients are taken into account. The statistical analysis carried out when comparing this estimator with two others also built in this work and others available from the literature, revealed its superior performance. The validating test by using the spectral classification from the Montreal White Dwarf Database showed an accuracy larger than 90\%, together with a precision and a sensitivity of 94\%. 

The excellent results in the identification of DA and non-DA white dwarfs led us to apply this estimator to the entire population with an effective temperature above 5500\,K. A total of 8150 objects (representing the 99\% of objects of our catalogue in that range of temperatures) have been spectrally classified in these two main groups. This classification has allowed to precisely determine the proportion of DA and non-DA objects in the different regions of the {\it Gaia} HR diagram.

Our results showed that the A branch in the region 0.1\,$<$\,$G_{\rm BP}-G_{\rm RP}$\,$<$\,0.5\,mag is practically formed by DA stars (94\%), while the B branch, although mainly constituted by non-DA white dwarfs (65\%), contains a significant fraction of DA white dwarfs (35\%). These results imply, on one hand, the confirmation of a bimodality in the DA mass distribution. Some hypothesis have been outlined such a stepper initial-to-final mass relation, or the contribution of white dwarf mergers \citep[e.g.][]{ElBadry2018,Kilic2018}. On the other hand, the practically nonexistence of non-DA white dwarfs in the A branch is indicative that helium-pure atmospheres suffer from some mixing process, leading to mixed envelopes for temperatures below 12\,000\,K. However, we cannot ensure that all non-DA white dwarfs in the B branch present a mixed atmosphere. Assuming the bimodality argument of the DA population caused by the flattened initial-to-final mass relation, in the case of non-DA white dwarfs, this scenario would imply that 20\% of non-DAs are genuine $\sim$\,0.8\,\msun\ white dwarfs. In any case, the ultimate characterization of these atmospheres is beyond the scope of the present work.

Our analysis also allowed us to derive stellar parameters by means of photometric interpolation in updated cooling sequences from La Plata models with hydrogen-pure or helium-pure atmosphere models. This way we built, for instance, the mass distribution for DA and non-DA stars. Our results showed a bimodality in the DA mass distribution, as previously commented and already reported in the literature, with a main peak at $\sim$\,0.58\,\msun\ and a secondary one at $\sim$\,0.8\,\msun. An extended tail for white dwarfs more massive than 0.8\,\msun\ is also observed for $T_{\rm eff}$\,$>$\,12\,000\,K, that disappears for lower temperatures due to a combination of crystallization and other physical process along with incompleteness bias effects. It was also noticed that for temperatures below 7000\,K the second peak fades. A possible cause is that, for this low range of temperatures, crystallization starts to happen for white dwarf less massive than 0.8\,\msun, smoothing out the mass distribution between the two peaks. Regarding the non-DA mass distribution, although a preliminary guess was done by using helium-pure models, some conclusions may be drawn. A clear peak at $\sim$\,0.58\,\msun\ mainly formed by genuine DB stars is present for temperatures above 12\,000\,K. This peak is moved to $\sim$\,0.72\,\msun\ at lower temperatures as a consequence of fitting the B branch with helium pure atmospheric models. However, for the lowest temperatures, the initial peak at even slightly less massive value $\sim$\,0.56\,\msun\ reappears. These facts indicate that individual spectral identification of white dwarfs atmosphere -- i.e., DB, DBA, DQ, DZ, etc -- is required in other to properly determine the non-DA mass distribution.

Finally, our analysis also allowed us to derive a precise ratio of non-DA to DA white dwarfs as a function of the effective temperature. Our spectral evolution distribution revealed a $\sim$\,13\% non-DA fraction nearly-constant smoothly increasing up to effective temperatures of $\sim$\,13\,000\,K, in excellent agreement with most of the recent published distributions. For lower temperatures, a marked growth was found reaching its maximum fraction of 36\% at $\sim$\,10\,500\,K, in perfect agreement with the spectral evolution function of \citet{Cunningham2020} and also compatible with that from \citet{Ourique2020}. Although for lower temperatures our result should be taken with caution, we found a decrease in the ratio of non-DA leading to a fraction of $\sim$\,31\% at $\sim$\,5500\,K.

In summary, the conjunction of accurate white dwarf models, the excellent observational database provided by {\it Gaia}, and the capabilities of automated tools such as the Virtual Observatory SED Analyser, allowed us to achieve the up-to-date largest spectral characterisation of the volume-limited white dwarf population in our solar neighbourhood, open thus the door to future studies such as its luminosity function or the star history of this region of the Galaxy.

\section{Data availability}
\label{results}

The {\it 100\,pc White Dwarf Gaia-DR3 Catalogue} is available online as supplementary material hosted by the jornal and at {\it The SVO archive of {\it Gaia} white dwarfs}\footnote{\url{http://svo2.cab.inta-csic.es/vocats/v2/wdw}} (see Appendix~\ref{append}) at the Spanish Virtual Observatory portal\footnote{\url{https://svo.cab.inta-csic.es/docs/index.php?pagename=Archives}}.

\section*{Acknowledgements}

We are greatly indebted to Detlev Koester for providing us with his white dwarf model atmosphere spectra. S.T. acknowledges fruitful discussion and data provided by Carlos L\'opez-Sanjuan. We also thank to the referee Dr Jay Holberg for his valuable comments.

F.J.E. acknowledges financial support from the Servicio Público de Empleo Estatal, Spain. P.C. acknowledges financial support from the Government of Comunidad Autónoma de Madrid (Spain) via postdoctoral grant `Atracción de Talento Investigador' 2019-T2/TIC-14760. ST and ARM acknowledge support from MINECO under the PID2020-117252GB-I00 grant. ARM acknowledges support from Grant RYC-2016-20254 funded by MCIN/AEI/10.13039/501100011033 and by ESF Investing in your future. R.M.O. is funded by INTA through grant PRE-OBSERVATORIO. M.E.C. acknowledges support from NASA grants (HGC) 80NSSC17K0008 and (LWS) 80NSSC20K0193 and the University of Colorado Boulder.

This work has made use of data from the European Space Agency (ESA) mission {\it Gaia} (\url{https://www.cosmos.esa.int/gaia}), processed by the {\it Gaia} Data Processing and Analysis Consortium (DPAC, \url{https://www.cosmos.esa.int/web/gaia/dpac/consortium}). Funding for the DPAC has been provided by national institutions, in particular the institutions participating in the {\it Gaia} Multilateral Agreement. This work has made use of the Python package {\it GaiaXPy}, developed and maintained by members of the {\it Gaia} Data Processing and Analysis Consortium (DPAC) and in particular, Coordination Unit 5 (CU5), and the Data Processing Centre located at the Institute of Astronomy, Cambridge, UK (DPCI). This publication makes use of VOSA, developed under the Spanish Virtual Observatory (https://svo.cab.inta-csic.es) project funded by MCIN/AEI/10.13039/501100011033/ through grant PID2020-112949GB-I00. We extensively made used of Topcat \citep{Taylor05}. This research has made use of the VizieR catalogue access tool, CDS, Strasbourg, France. We acknowledge use of the ADS bibliographic services.

\bibliographystyle{mnras}
\bibliography{references} 


\appendix

\section{Online catalogue service}
\label{append}

In order to help the astronomical community on using our catalogue of white dwarfs, we developed a wed archive system that can be accessed from a webpage\footnote{\url{http://svo2.cab.inta-csic.es/vocats/v2/wdw/}} or through a Virtual Observatory ConeSearch\footnote{e.g. http://svo2.cab.inta-csic.es/vocats/v2/wdw/cs.php?RA=301.708\&DEC=-67.482\&SR=0.1\&VERB=2}.

The archive system implements a very simple search interface that permits queries by coordinates and radius as well as by other parameters of interest. The user can also select the maximum number of sources (with values from 10 to unlimited) and the number of columns to return (minimum, default, or maximum verbosity).

The result of the query is a HTML table with all the sources found in the archive fulfilling the search criteria. The result can also be downloaded as a VOTable or a CSV file. Detailed information on the output fields can be obtained placing the mouse over the question mark (``?") located close to the name of the column. The archive also implements the SAMP\footnote{\url{http://www.ivoa.net/documents/SAMP/}} (Simple Application Messaging) Virtual Observatory protocol. SAMP allows Virtual Observatory applications to communicate with each other in a seamless and transparent manner for the user. This way, the results of a query can be easily transferred to other VO applications, such as, for instance, TOPCAT.


\bsp	
\label{lastpage}
\end{document}